\documentclass[3p,times]{elsarticle}

\usepackage{epsfig}
\usepackage{graphicx}

\usepackage{amssymb}

\usepackage{lineno}

\usepackage[figuresright]{rotating}

\usepackage{amssymb}
\usepackage{subfigure}
\begin{document}
%

\begin{frontmatter}
%
%
%
%
%
%
\title{Dynamic Fuzzy c-Means (dFCM) Clustering and its Application to Calorimetric Data Reconstruction in High Energy Physics}
%
%

\author{Radha Pyari Sandhir, Sanjib Muhuri and Tapan K. Nayak}

\address{Variable Energy Cyclotron Centre, Kolkata - 700064, India}

\begin{abstract}
In high energy physics experiments, calorimetric data reconstruction requires a suitable clustering technique in order to 
obtain accurate information about the shower characteristics such as position of the shower and energy deposition. Fuzzy 
clustering techniques have high potential in this regard, as they assign data points to more than one cluster, 
thereby acting as a tool to distinguish between  overlapping clusters. 
Fuzzy c-means (FCM) is one such clustering technique that can be
applied to calorimetric data reconstruction. However, it has a
drawback: it cannot easily identify and distinguish  clusters that are
not uniformly spread.  
A version of the FCM algorithm called dynamic fuzzy c-means (dFCM) allows clusters to be generated and eliminated as required, with
the ability to resolve non-uniformly distributed clusters. Both the FCM and dFCM algorithms have been studied and successfully applied to
simulated data of a sampling tungsten-silicon calorimeter. It is seen that the
FCM technique works reasonably well, and at the same time, the use of the dFCM technique
improves the performance.

\end{abstract}

\end{frontmatter}

\section{Introduction}
\label{}
The human mind can easily grasp concepts like imprecision,
uncertainty, partial truth and approximation. Conventional (hard)
computing, i.e., computing with traditional electronic
computations has limitations, as it only sees in terms of `black' and
`white', or more accurately, zeroes and ones.  Though this way of
computing is sufficient for a large number of tasks, it cannot handle a problem
for which there is not yet enough information to calculate
definitively what the answer is.  Soft computing, on the other hand,
uses nature as a role model, allowing computers to take
uncertainty and imprecision into account, in much the same way the
human mind does.  
Soft computing surfaced as a formal 
computer science area of study~\cite{Zadeh} in the early 1990's, and acts as the
emerging field of `computational intelligence' 
essentially adding `intelligence' to computing techniques~\cite{andries} .  It is an
essential tool for a good decision making model, 
and has been used in a wide variety of disciplines from bioinformatics
to aeronautical engineering to image processing. 
Computational intelligence has also found its way to detector physics,
especially for high energy physics experiments, and has been used 
to determine detector performances~\cite{SPalViyogiFCM,Viyogi,ambriola}, as
well as 
extract physical information of interacting particles~\cite{whiteson,muller,cdf,Yu}. 

In high energy physics experiments, one of the essential steps in the extraction of physical information
from particle detectors is to reconstruct the characteristics of the
incoming particles. 
A calorimeter is normally used for
accurate characterization of an incoming particle in terms of its position and energy.
Depending on the structure, calorimeters can be categorized into two types: 
homogeneous and sampling. A homogeneous calorimeter is made up of a single 
block of material that acts as an absorber as well as an active medium from which the signal can be collected. 
On the other hand, sampling calorimeters are segmented in the
longitudinal as well as in the transverse directions, consisting of layers of
absorber and active detector combinations. Each segment or cell acts as an
independent detector. To reconstruct the physical information one
needs to identify the group of hit cells associated with an incoming
particle, and determine the most probable position of the particle and
its energy.  This is done by the use of a clustering
algorithm. Additionally, the clustering techniques can also be used to
classify different sets of events~\cite{Mjahed}.

A number of clustering algorithms exist in the literature, and can be
employed for high energy physics data reconstruction. 
Algorithms like the contiguity method and cellular automata, search
neighborhoods of cells would not be as 
useful for distinguishing between overlapping clusters found in high
density environments.  Some other methods, like 
deterministic annealing, require
parameters that depend very strongly on the data pattern, and would
not serve well as a generic clustering algorithm.  
Hard clustering techniques such as local maxima search, connected-cell search and k-means clustering simply assign a data
point to a cluster. A data point either lies in a cluster or it does
not, and so, overlapping clusters are hardly distinguishable. 
In \emph{fuzzy} clustering, on the other hand, data points are
represented by degrees of membership, 
i.e., they lie within clusters to varying
degrees. The term `fuzzy' is used because an
observation may in fact lie in more than one cluster simultaneously, as is the case
with many high energy physics applications.  
In this article, a type of fuzzy clustering called fuzzy c-means (FCM)~\cite{nock,Bezdekbook, Bezdekvalidity}
is studied and applied to simulated data of a sampling
tungsten-silicon calorimeter.
A modification to the FCM algorithm~\cite{dFCM}, that allows the clustering to be
carried out dynamically was found to resolve particle clusters well specifically in the case of nonuniformly distributed clusters. 

This article is organized as follows.  
Section~\ref{sect:FCM} discusses the FCM algorithm including possible  
modifications. One such modification is the dynamic FCM (dFCM) technique, which is discussed in Section~\ref{sect:dFCM}.
Details of the calorimetric configuration are outlined in 
Section~\ref{sect:Configuration}.
The power of FCM to
distinguish between overlapping clusters is demonstrated in
Section~\ref{sect:FCMapplication} by performing clustering on two
photon clusters obtained by neutral pions of varying energies. The
ability of dFCM to resolve non-uniformly distributed clusters with
ease unlike FCM, is demonstrated in
Section~\ref{sect:dFCMapplication}, followed by its application to
calorimetric data.  The paper concludes in
Section~\ref{sect:summary} with a summary and a
remark on future perspectives.

\section{Fuzzy Clustering and the FCM Algorithm}\label{sect:FCM}

Fuzzy clustering is a technique in which the allocation of data
points to clusters follows fuzzy logic, thereby providing the means to separate overlapping
clusters.
Fuzzy clustering is appropriate for applications in detectors,
as showers generated by particles may
have profiles that are continuous or overlapping on the detector
planes. Fuzzy c-means (FCM) is a commonly used fuzzy clustering 
algorithm~\cite{nock,Bezdekbook, Bezdekvalidity}. 
It is a version of the k-means algorithm that
incorporates fuzzy logic, so that each point has a weak or strong
association to the cluster, determined by the inverse distance to
the center of the cluster.  The centers obtained
using the FCM algorithm are based on the geometric locations of the
data points. 
The FCM algorithm has been used in \cite{Suliman} to track gamma
rays in segmented detectors, and in \cite{SPalViyogiFCM} to find
clusters in the preshower detector in high energy heavy ion experiments.

For a set of data points $X$, FCM seeks to minimize the  objective function -- the
weighted within groups sum of squared errors $J_{m}$:
\begin{equation}
J_{m}(U,V;X)=\sum_{k=1}^{n}\sum_{i=1}^{C}(u_{ik})^{m}\|x_{k}-v_{i}\|^{2},
\end{equation}
where $V=(v_{1},v_{2},\ldots,v_{C})$ is a vector of unknown cluster
centers or centers, $U$ consists of the memberships $u_{ik}$ of
the $k^{th}$ point in the $i^{th}$ cluster, and
$\|x\|=\sqrt{x^{T}x}$ is any inner product norm.
The fuzzy factor $m$ normalizes and fuzzifies the memberships so that their sum is 1.  A value of 2 implies linear normalization, whereas when $m$ is closer to 1, the cluster center closest to the point is given much more weight than the others, making the algorithm similar to k-means.
Optimization of $J_{m}$ is based on iteration through certain
necessary conditions. Following the {\it FCM Theorem}~\cite{Bezdekbook}
If $D_{ik}=\|x_{k}-v_{i}\|>0$ for all $i$ and $k$, then (U,V) may
minimize $J_{m}$ only if, when $m>1$,
\begin{equation}
u_{ik}= \bigg [ \sum_{j=1}^{C}\bigg (\frac{D_{ik}}{D_{jk}} \bigg
)^{\frac{2}{m-1}}\bigg ]^{-1} \label{FCMmemb}
\end{equation}
where $1\leq i \leq C$, ~~ $1\leq k \leq n$, ~~and
\begin{equation}
v_{i}=\frac{\sum_{k=1}^{n}(u_{ik})^{m}x_{k}}{\sum_{k=1}^{n}(u_{ik})^{m}}.
\label{FCMcenters}
\end{equation}

\begin{figure}[hbtp]
\begin{center}
\includegraphics[scale=.4]{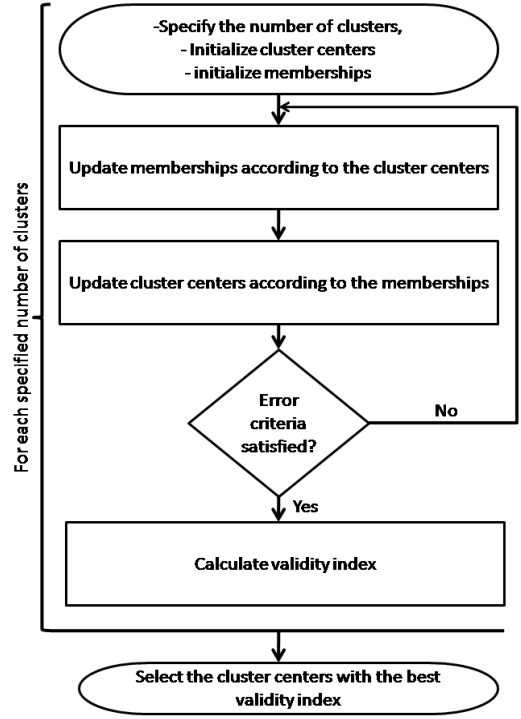}
\caption{The flow chart for the fuzzy c-means (FCM) algorithm.} 
\label{fig:FCMflowchart}
\end{center}
\end{figure}

Alternating optimization (AO) is the iteration technique that is
most often used in this algorithm. It simply loops through one cycle
of estimates for $V_{t-1} \rightarrow U_{t} \rightarrow V_{t}$ until
some error criteria is reached.  An error threshold $\epsilon$ can
be specified so that the error criteria is $\|V_{t-1}-V_{t}\|_{err}
\leq \epsilon$. Defuzzification, that is, determining which point lies in which
cluster, can be done by looking at the memberships of a 
point associated with each cluster.  A point belongs to a cluster if
its corresponding 
membership is the maximum out of the point's memberships in all of the clusters.
Figure~\ref{fig:FCMflowchart} shows the flow chart of the FCM
algorithm. The algorithm runs for a specified 
number of clusters.  This number can be varied, and the best set of
clusters can be 
selected by means of a validity index.

\subsection{Validity Indices}\label{sect:valindex}

A \emph{validity index} seeks to determine how well the data  is represented by the selected clusters.  
There are a number of validity indices available in the
literature~\cite{Bezdekvalidity}. 
The  Xie-Beni index \cite{XB} is one such validity index that is
widely used 
because of its dependence on both memberships as well as geometric
distances. More explicitly, the index depends on the distances between data points and the centers of
clusters as well as the distance between cluster centers. 
Other indices are based on other aspects of the data.  For instance,
the \emph{partition coefficient} depends 
only on the memberships of the data:
\begin{equation} \label{eq:PC}
PC(c)=\frac{1}{n} \sum_{i=1}^{c} \sum_{j=1}^{n} \mu_{ij}^{2},
\end{equation}
where $n$ is the number of data points, $c$ is the number of clusters, and $\mu_{ij}$ is the membership of the $j^{th}$ data point in the $j^{th}$ cluster.  It varies from $1/c$ to 1.  This index is to be maximized.
In order to remove the dependence of $PC$ on $c$, the Modified
Partition Coefficient was defined as \cite{DaveMPC}:
\begin{equation}\label{eq:MPC}
MPC(c)=1- \frac{c}{c-1}(1-PC),
\end{equation}
which varies from 0 to 1.  This index is to be maximized as well.
The Xie-Beni
index is chosen for the present study,
 though it may be noted that any validity index preferred by the experimenter may be employed.
The Xie-Beni index
$\upsilon_{XB}$ is defined as follows:
\begin{equation}
\upsilon_{XB}(U,V;X)=\frac{\sum_{i=1}^{C}\sum_{k=1}^{n}u_{ik}^{2}
\|x_{k}-v_{i} \|^{2}}{n(\underbrace{\min}_{i\neq j} \{ \|v_{i} -
v_{j} \|\})}.
\end{equation}
It is essentially the ratio of the total variation of the cluster
centers and memberships of the observations in the groupings to
the separation between the cluster centers, and minimization of the index leads to better clusters.
The larger the separation between clusters and the more
closely packed the points in the cluster, the better the clustering.  This index has no upper bound.

\subsection{Modifications to the FCM Algorithm} \label{sect:Mod}

Over the years, a number of application-specific modifications have
been made to the FCM algorithm.
As FCM clustering is based on the Euclidean distance between data
samples,
 it gives each data point and each dimension (or feature) the same
 importance. 
A modification to FCM using feature-weight learning was looked 
into \cite{SimilarityClustering,FWLearning}, however, in the present study, each of the three 
dimensions considered in the clustering should be given equal importance.
The drawbacks of FCM include the dependence on the fuzzy factor $m$, which may vary from one data set to other~\cite{FCMParameters}, and the fact that
it treats outliers in the same way it treats data points lying in the bulk of
the data.  In order to address these concerns, the suppressed FCM
algorithm \cite{SFCM} was looked into.  This algorithm prizes the
biggest memberships and suppresses the others with a weighting
factor.

A modified version of the FCM algorithm \cite{dFCM} has been developed,
which has the main advantage that it dynamically finds clusters as data
streams in, deleting and generating clusters as needed. The decision
making for valid clusters is made through the use of a
validity index. Theoretically, the method does not require a maximum
number of clusters, only a minimum, which gives it an edge over the
energy-weighted modification.  It also adapts to the data pattern at
each instant. This modified algorithm is called dynamic Fuzzy c-Means
(dFCM), which is discussed in detail in the next section.

\section{The dynamic Fuzzy c-Means (dFCM) Algorithm}
\label{sect:dFCM}

\begin{figure}[h]
\begin{center}
\includegraphics[scale=.34]{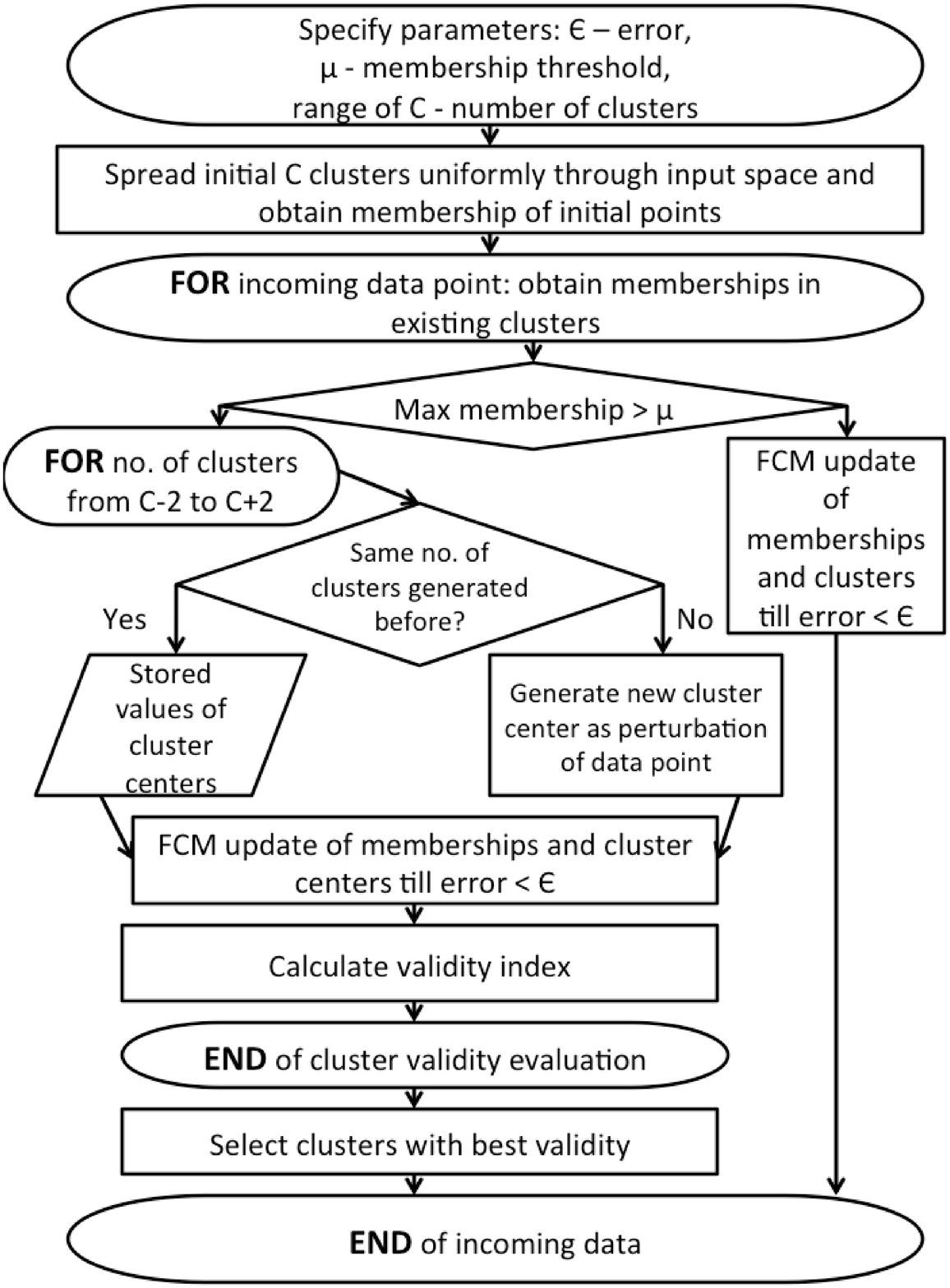}
\caption{The flow chart of the dynamic FCM (dFCM) algorithm.}
\label{fig:DFCMflowchart}
\end{center}
\end{figure}

The dFCM algorithm \cite{dFCM} is a modification of the fuzzy
c-means algorithm, allowing cluster centers to be adaptively
updated as data points keep streaming in.  If a new cluster is
formed, then a new cluster center is automatically generated.
Figure~\ref{fig:DFCMflowchart} gives a detailed flow chart of the algorithm.
The working of the algorithm is as follows:

\begin{enumerate}
\item  To start with, we assume that we already have a few of the incoming data points
  at hand.  From these data points, we can roughly estimate - but not restrict
ourselves to - the range of the incoming data.  Initially, a few parameters are
 specified, namely, the membership threshold, $\mu$, the FCM
error criteria $\epsilon$, and the bounds of the variable $C$ -- the number of clusters. The initial number of points is taken to be low, just to kick-start the clustering and give the algorithm a brief idea about the values it is dealing with.  Unless the initial number of points is so large that hints of clusters can already be seen, the clustering wonÕt be affected.  

\item If the minimum value of the variable $C$ is $C_{min}$ then,
  initially, $C_{min}$ 
cluster centers are generated uniformly within the input space.  
Note that this number must be greater than or equal to 2 as the
 unit cluster is not allowed in the FCM algorithm. 
Once the initial cluster centers have been specified, 
the memberships of the initial data points are found using Equation~\ref{FCMmemb}.

\item The data points are now allowed to stream in, one by one.  When
  a new data point arrives, 
its memberships in the clusters present are calculated.  If the
maximum 
membership value associated with this point is greater than or equal
to the membership 
threshold $\mu$, then a simple AO update takes place, 
as outlined in Section~\ref{sect:FCM}.  
This means that the data point belongs to at least one of the clusters to an extent that matches or exceeds $\mu$.

\item Let $C$ be the number of clusters present at a given time. If
  the maximum membership 
of the data point falls \emph{below} $\mu$, then the validity of the
present 
number of cluster centers $C$ is compared with each of the
validities of $L= C-2$ to $C+2$ 
clusters in the following manner:
\begin{itemize}
\item The stored values of the cluster centers are checked to see
  if $L$ clusters have been 
generated and updated at a previous time. If so, then the old values
are updated using the 
FCM algorithm, and the validity index is evaluated.
\item If $L$ cluster centers have not been generated before, then
  the stored values of $L-1$ 
centers are used, and an $L^{th}$ center is generated by
slightly perturbing the data point. 
The AO update is carried out, and the validity index is evaluated.
\item The cluster centers that generate the best value of the validity index are taken.
\end{itemize}
\item The process continues until the data points stop streaming in.
\end{enumerate}

\subsection{The Membership Threshold ($\mu$):} 
\label{condition}

The purpose of the membership threshold, $\mu$, is to avoid
evaluating cluster validity \emph{each} time a data point comes in.
That is, if the data point lies to a specified satisfactory extent
in a cluster, then it is not necessary to check if other clusters are
better. However, the role of $\mu$ may be allowed to change, depending on
what the specific application of the dFCM algorithm may require. The
experimenter is free to set the conditions that need to be met in
order to evaluate cluster validity.  For instance, the condition
that is specified in this work is \emph{whether or not the data point
lies within a cluster to a satisfactory extent}.  In another
application, it may be that validity index evaluation need only take
place if the new updated centers are \emph{significantly
different} from the old ones.  The factor $\mu$ can then be used in
a condition that may look like:
\begin{equation}
\|V_{old}-V_{new}\| > \mu,
\end{equation}
i.e., if the distance between the two sets of coordinates is greater
than $\mu$, then validity is evaluated. In a situation like this,
the AO update would take place automatically once a data point
streams in, without any check on its membership. Once the centers
are updated, $\mu$ can be used to check whether they are
significantly different from the previous ones, and therefore
whether or not validity should be evaluated. This prevents redundant calculations.

\subsection{Cluster Centers and Validity}

Cluster centers are kept track of because validity indices may
change as data points keep streaming in.  At a certain point in
time, two cluster centers may be sufficient, whereas at a later
point, the algorithm may call for four cluster centers.  The reverse
may be true as well, and so, in order to keep the window open for 
cluster number possibilities, $C-2$ to $C+2$ clusters are checked, where
$C$ is the actual number of cluster centers at a particular point
in time. This window is flexible, and may be modified to be wider or
narrower, 
depending on the preference of the experimenter.
Note that the values of the cluster centers are made to
fall within the initially specified range.  However, the maximum
number of cluster centers need not be specified if the
experimenter prefers to keep the upper limit open.
Any validity index favored by the experimenter may be employed, as
the algorithm does not specifically depend on the type of validity
index used.

\subsection{Running Time of dFCM}

The running time for the dFCM algorithm coupled with the Xie-Beni index was calculated and found to be $O(an^{2})$, where $n$ is the size of the data, and $a$ is the maximum number of AO updates required, explained below.  The worst-case scenario was taken, in which a new cluster is generated for every incoming point, and there are no clusters to begin with.
The running time had been obtained as follows: Simple arithmetic and logical operations take constant amounts of time.  Unless a calculation that uses these operations explicitly depends on the size of the data or on a search, then the calculation also takes a constant amount of time.  As each data point streams in, a check is performed to see if it lies within the established clusters to a satisfactory extent (step 3).  Since we're assuming that each new point acts as a new cluster, each possible `check' with the present clusters (i.e. in this case, the data points themselves) takes place.  
When the first point streams in, no check is performed, as there are no data points present.  When the second point streams in, only 1 check is performed.  When the next point streams in, 2 checks are performed, and so on.   The total number of checks is represented by the following series:
\begin{equation}\label{eq:series}
1+2+3+ \cdots + n-1 = \frac{n(n-1)}{2}.
\end{equation}
This implies a running time of order $O(n^{2})$ for the various
checks.  When a new cluster is generated, validity indices are
calculated for $L=C-2$ to $C+2$, and an AO update is performed (step~4).  

The running time of the update depends on the number of data points already present, say $k$, and the number of iterations required before the error criteria is reached, say $a_{k+1}$, as it may differ for each case.  The subscript $k+1$ indicates that the $(k+1)^{th}$ data point is streaming in. Therefore, the running time of the update can be considered to be $O(ka_{k+1})$. The running time of the validity index calculations can be taken as a constant ($l=C+2-(C-2)=4$) times $k$, as the validity index evaluation sums over all the data points present at the given time only once, and is performed an $l$ number of times.  
The total running time associated with each number contributing to Equation~\ref{eq:series} can be thought of as: $O(ka_{k+1})+O(lk)$ of which $O(ka_{k+1})$ dominates.  Since the order of the total running time of the AO updates can be thought of as:
\begin{equation}
1\cdot a_{2} + 2 \cdot a_{3}+ 3\cdot a_{4}+\cdots +(n-1)\cdot a_{n},
\end{equation}
where we define, $a=\max(a_{2},a_{3},\cdots ,a_{n})$.  Therefore, the total running time of the dFCM algorithm can be considered to be: $O(an^{2})$.

\subsection{Applications of dFCM}

The dynamic fuzzy c-means clustering technique can be applied in situations that involve online analysis of streaming data, in which adaptive information is required, or in which the data to be clustered is not uniform.  Most dynamic versions of available clustering techniques are application specific \cite{Ming,Ensan}.  However, dFCM is a generic algorithm that can be applied to a number of different situations.  For instance, \cite{dFCM} discusses a potential application as an adaptive rule extraction technique for fuzzy associative memories, in the field of soft computing.  In this paper, dFCM has been applied to high energy particle physics data reconstruction.  Another potential application may lie in time-series analysis and prediction, or updating databases.

\section{Calorimeter Concept and Design}
\label{sect:Configuration}

\begin{figure}[htbp]
\begin{center}
\includegraphics[scale=.4]{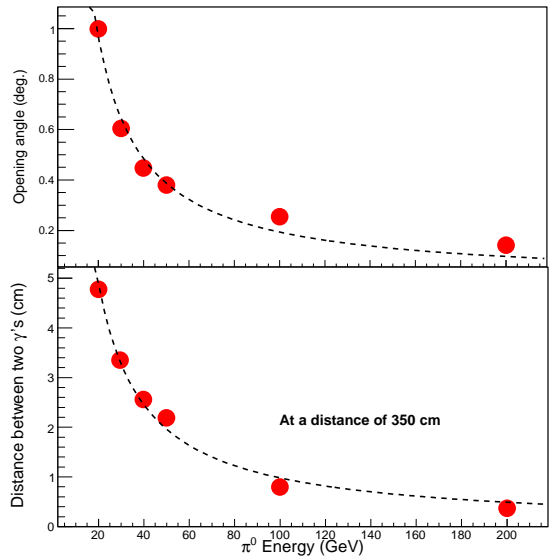}
\caption{
Opening angle (top panel) between two photons emitted from the decay of 
$\pi^0$ as a function of $\pi^0$ energy. 
The minimum distance between two photons at a distance of 350~cm from
the vertex is shown in the bottom panel. As the $\pi^0$ energy
increases, both the opening angle of the two decaying photons and 
the distance between them decrease. The dashed curves are fits to the
data points, which agree with theoretically calculated values. 
}
\label{fig:pi0decay}
\end{center}
\end{figure}

As a demonstration of the applicability of the FCM and dFCM clustering
techniques, a sampling calorimeter consisting of tungsten and silicon
layers was simulated using the GEANT4 package~\cite{geant4-1,geant4-2}.
A similar tungsten-silicon calorimeter has been developed and used by
the CALICE Collaboration~\cite{calice1,calice2}.
A sampling calorimeter consists of several planes of absorber material along with active medium planes.
A design of a tungsten-silicon calorimeter was made
with 20 layers, where each layer consisted of a 3 mm thick tungsten plate followed
by a 0.3 mm thick silicon sensor.
Being a high-Z element, tungsten converts high energy photons or
electrons into electromagnetic showers. The majority of the photons that
are emitted in high energy collisions are decay photons from
$\pi^0$. One of the major goals is to reconstruct $\pi^0$ and their
energy from the measured photon showers. The decay angle of the two emitted
photons decreases with the increase of the energy of $\pi^0$. 
The distance between the two emitted photons has been calculated for a
placement of the calorimeter at a distance of 
350~cm from the interaction vertex, as shown in Figure~\ref{fig:pi0decay}. As the $\pi^0$ energy increases,
the distance between the two photons decreases. The reconstruction of
the photon showers needs to be accurate in order to obtain the shower
positions of the photons and deposited energy. In order to measure $\pi^0$
energy accurately, tracking of the shower in different layers
is needed. Therefore, the position resolution of the detectors in the
sensitive medium has to be suitable.

\begin{figure}[h]
\begin{center}
\includegraphics[scale=.5]{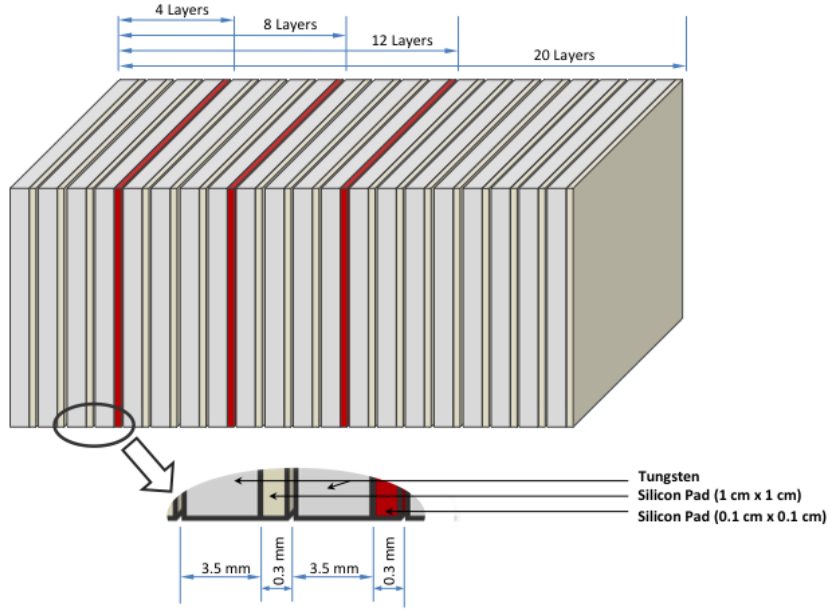}
\caption{The components of a tungsten-silicon sampling calorimeter consisting of 
20 layers of detectors. The three highlighted layers 
are made of highly granular 0.1~cm~$\times$~0.1~cm silicon pads, 
and the rest of the layers consist of 1 cm x 1 cm silicon pads.}
 \label{fig:calorimeter}
\end{center}
\end{figure}

\begin{figure}[htbp]
\begin{center}
\includegraphics[scale=.4]{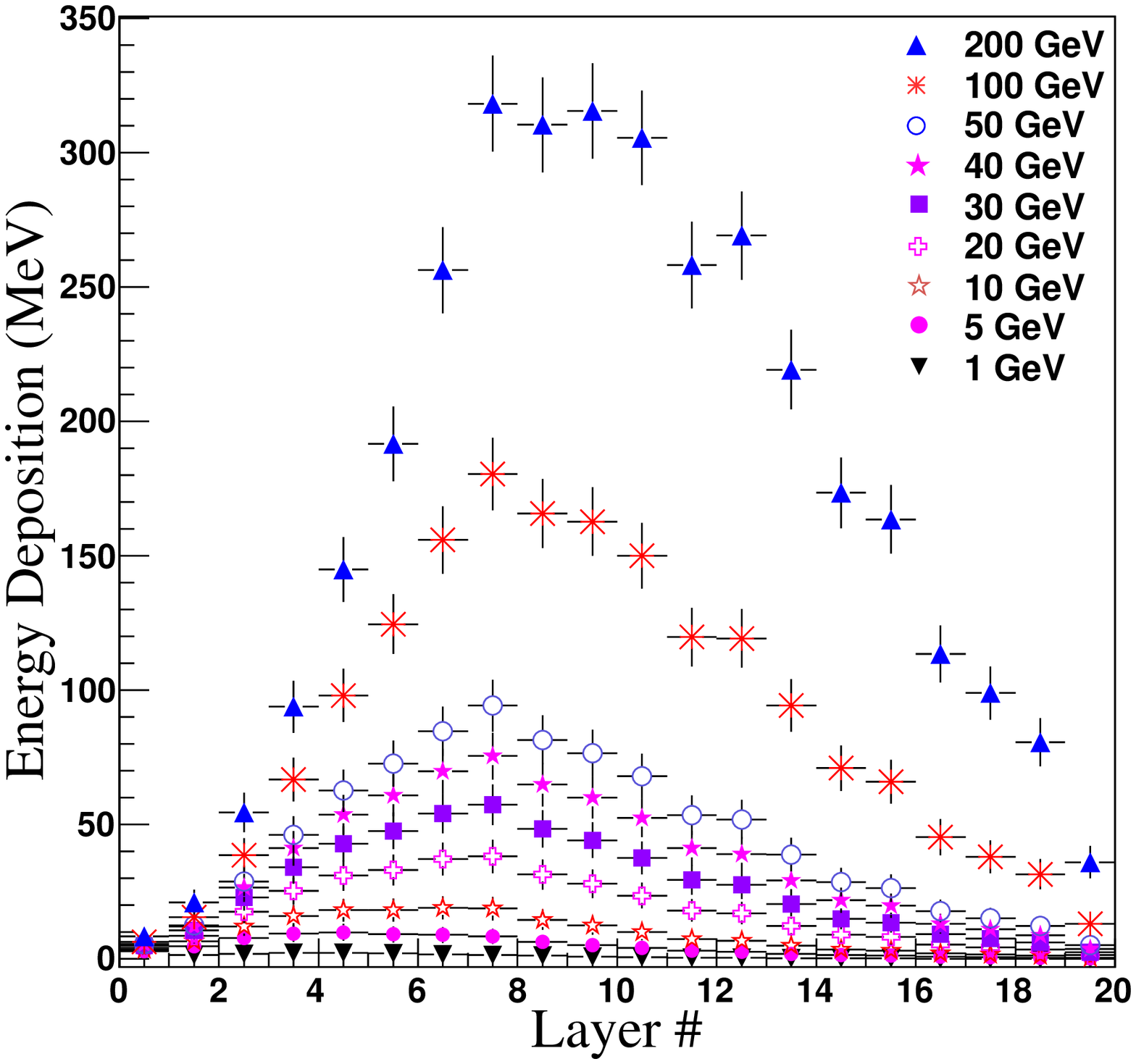}
\caption{
The longitudinal shower profile for energy deposition by photons of
different energies in various layers of the sampling calorimeter.
}
\label{fig:showerprofile}
\end{center}
\end{figure}

The calorimeter, 
shown in Figure~\ref{fig:calorimeter}, was designed keeping all of the requirements in
mind. The longitudinal shower profile
of the calorimeter, as shown in Figure~\ref{fig:showerprofile},
was studied in order to decide the granularity of the detector
layers. From this figure, it was found that the maximum of the shower
in terms of energy deposition occurs around  layer 4 for
photon energy of around 5 GeV, whereas for a photon of 50 GeV, the
shower maximum is around layer 8. 
Therefore, high granular planes were placed in the region of the shower maxima for accurate measurements.
For the purpose of simulation, only twenty layers were considered, 
out of which three layers (4,
8 and 12) were made of highly granular silicon pads, each with dimension of
0.1 cm $\times$ 0.1 cm.
The rest of the layers were made up of
1 cm $\times$ 1 cm silicon pad detectors. The three high granular planes
help to determine the shower position with high accuracy and thus help
in tracking the path of the incoming particles. 
The shower generation, energy deposition and the characteristics of
the shower produced by photons have been studied for the calorimeter
using both FCM and dFCM techniques.

\section{FCM applied to photon clusters}
\label{sect:FCMapplication}

\begin{figure}[htbp]
\begin{center}
\begin{minipage}{500pt}
\begin{tabular}{cc}
\includegraphics[width=225pt,height=180pt]{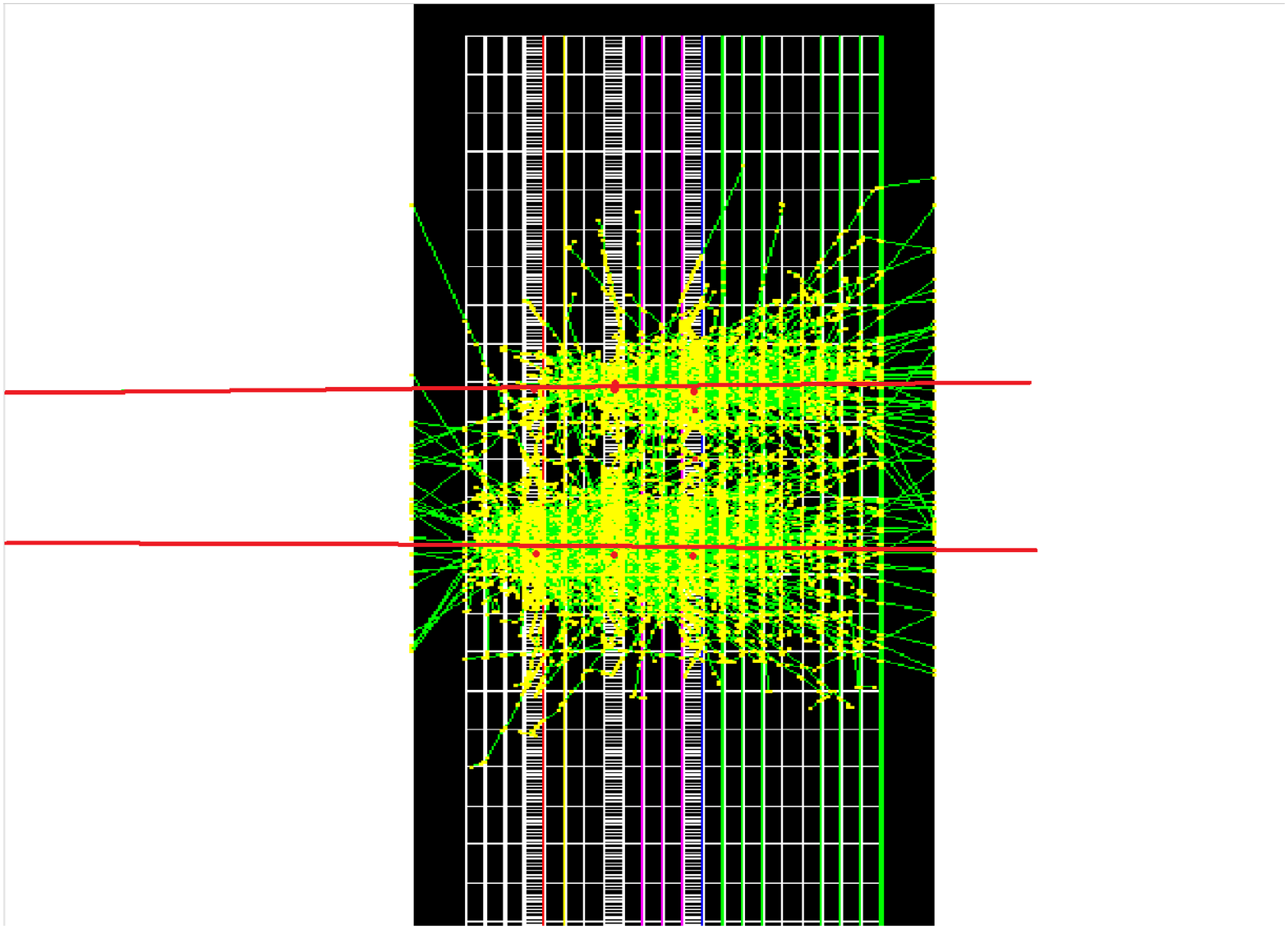} &
\includegraphics[width=205pt,height=180pt]{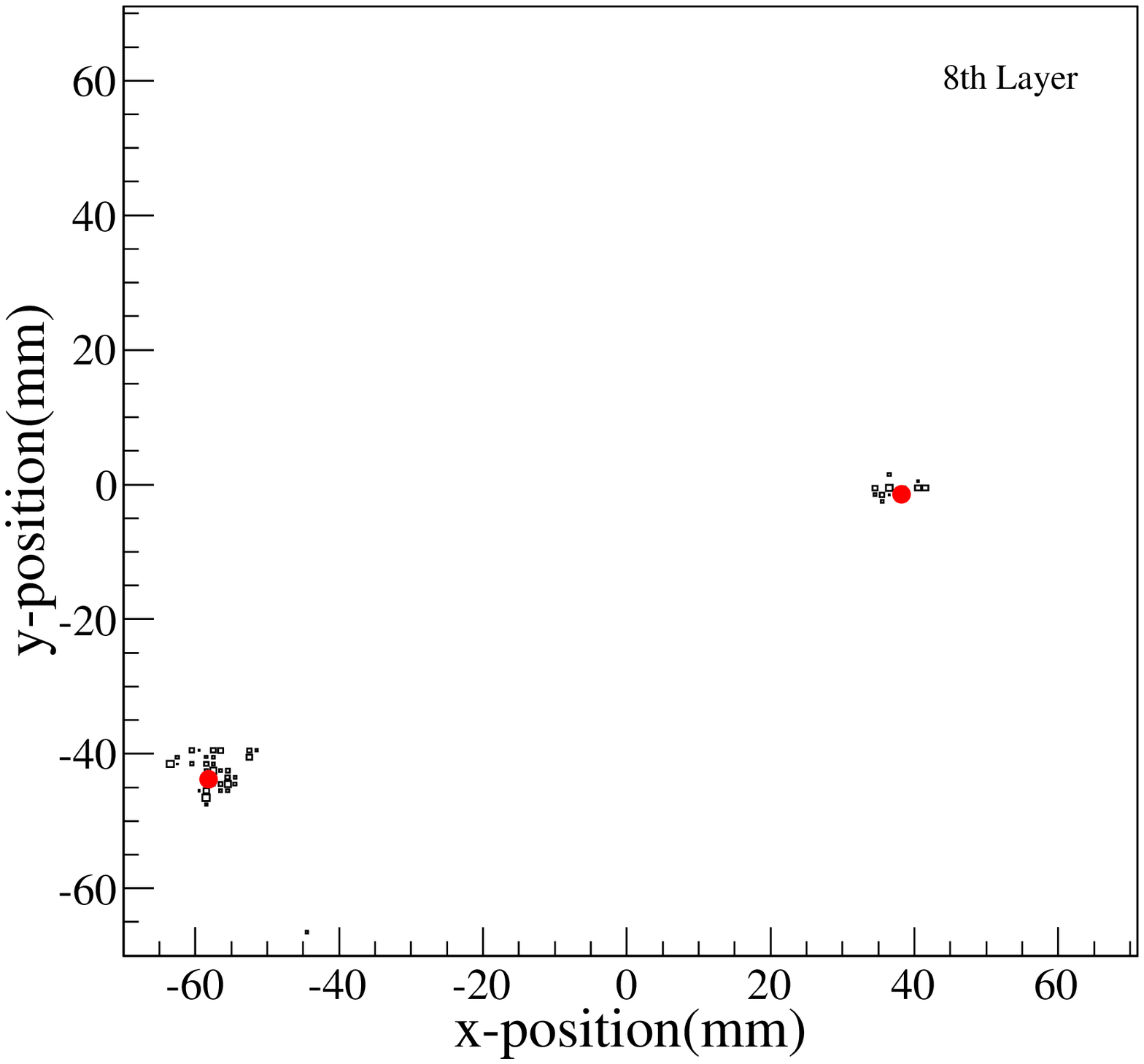} \\
(a) & (b)
\end{tabular}
\end{minipage}

\begin{minipage}{500pt}
\begin{tabular}{cc}
\includegraphics[width=232pt,height=180pt]{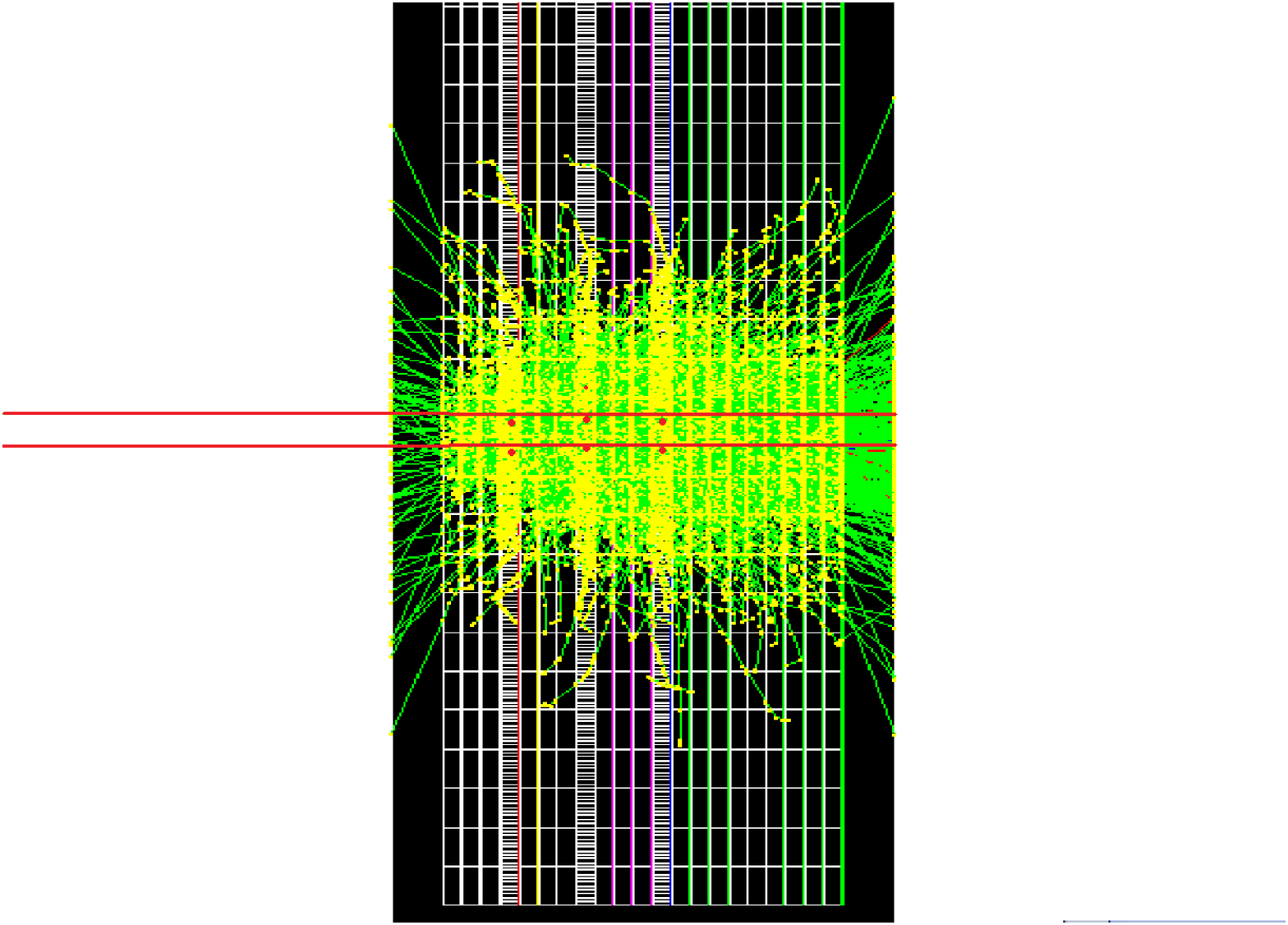} &
\includegraphics[width=200pt,height=180pt]{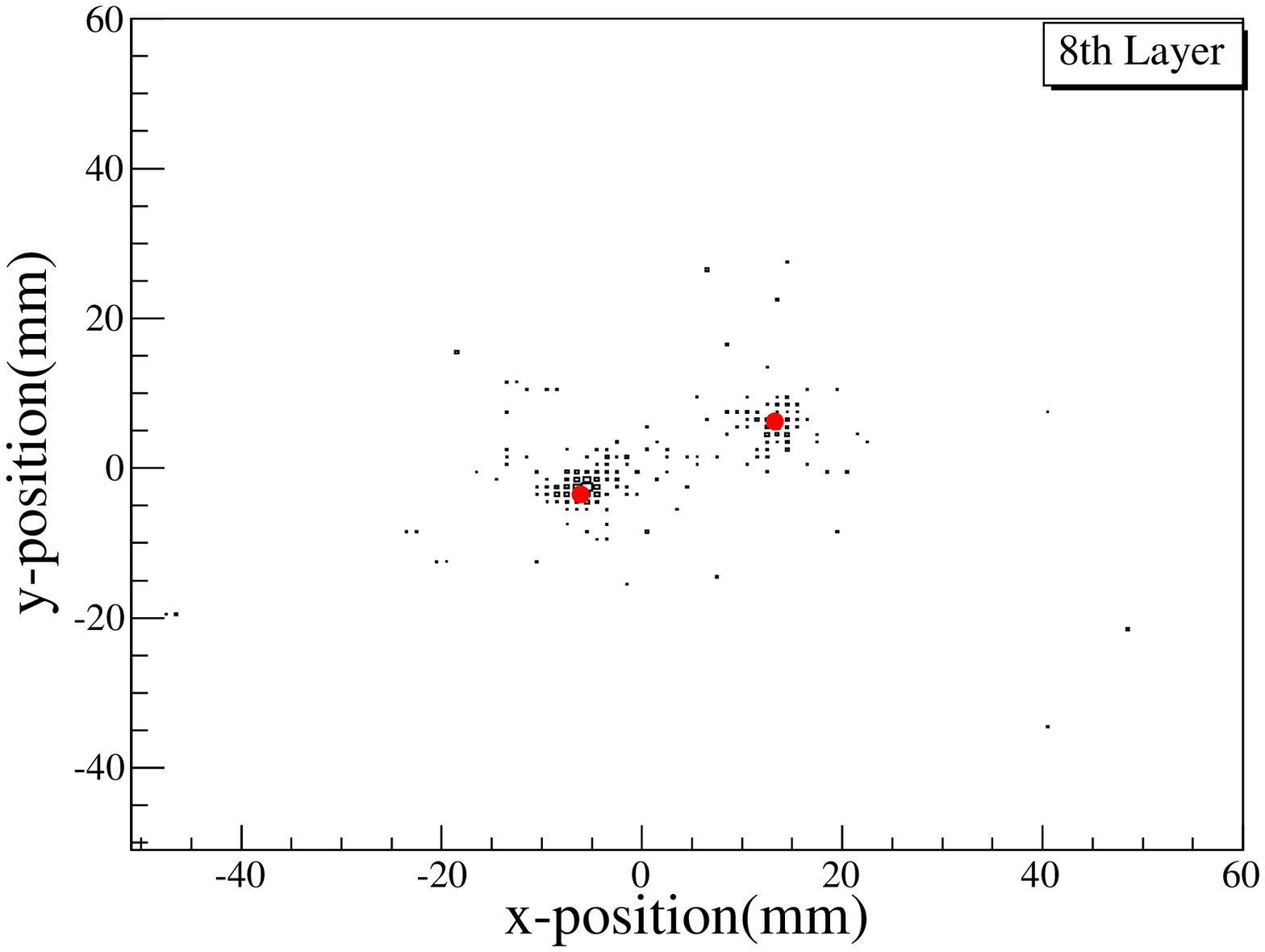} \\
(c) & (d)
\end{tabular}
\end{minipage}

\begin{minipage}{500pt}
\begin{tabular}{cc}
\hspace{0.1cm}\includegraphics[width=268pt,height=180pt]{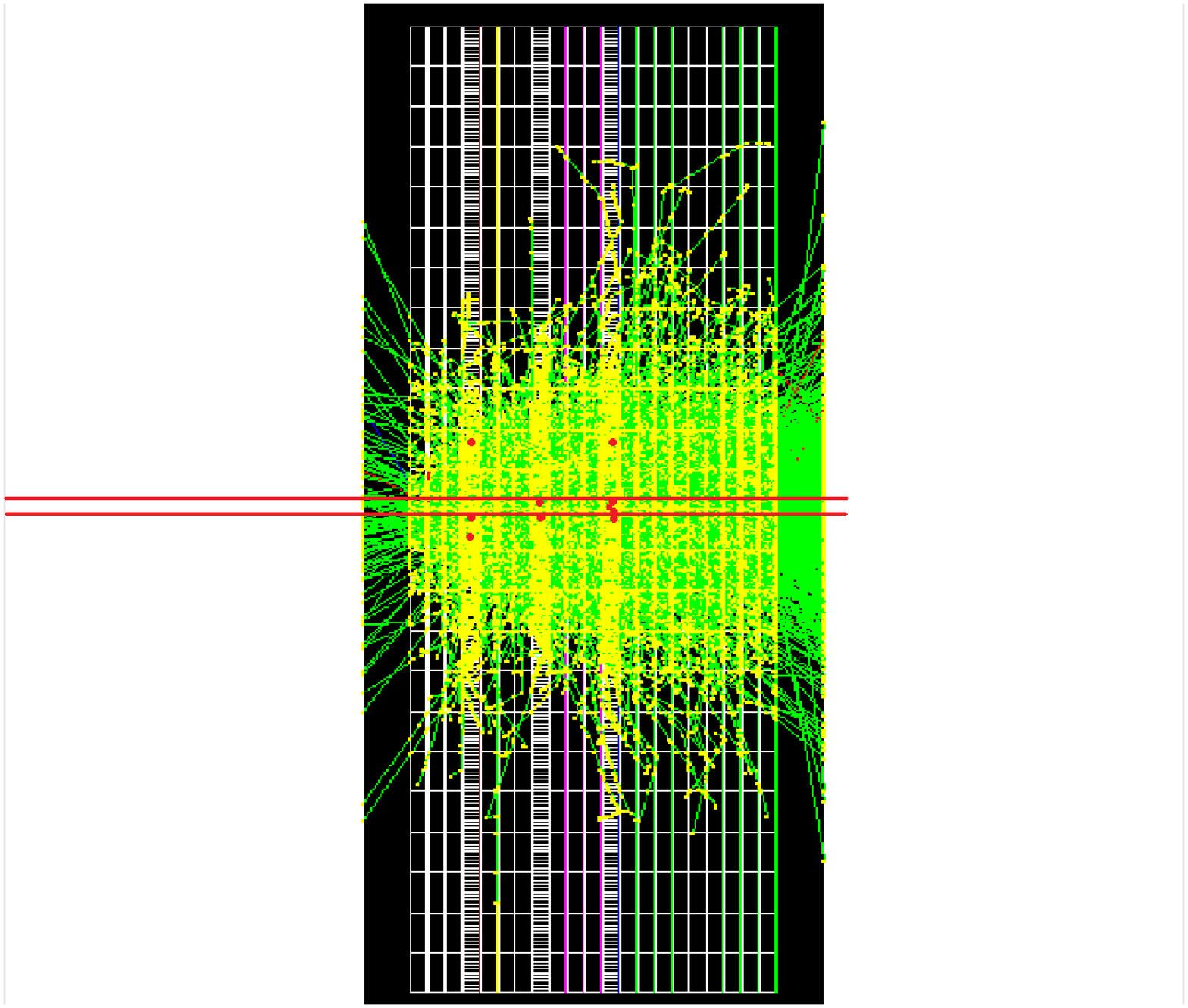} &
\hspace{-1.4cm}\includegraphics[width=205pt,height=180pt]{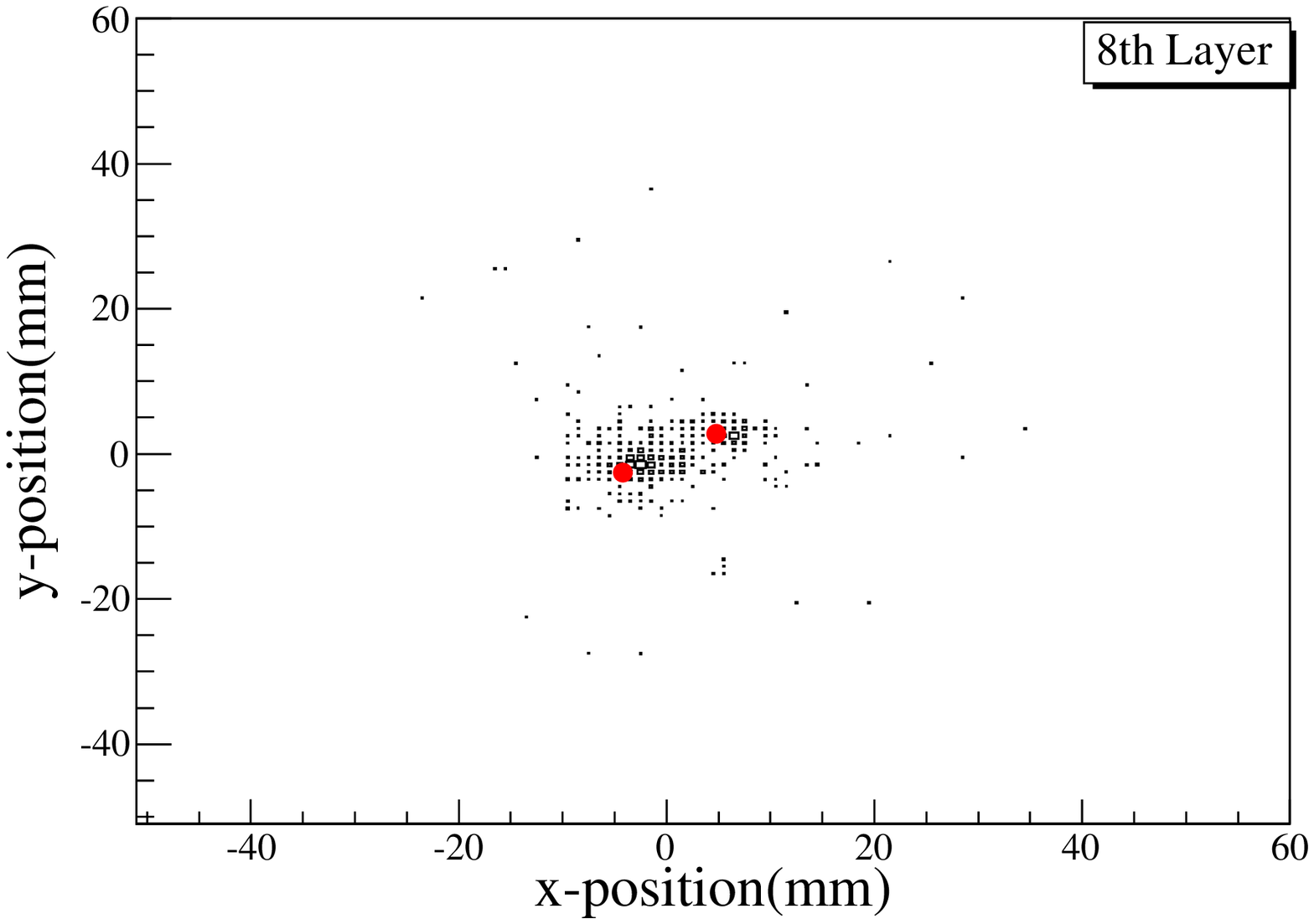} \\
(e) & (f)
\end{tabular}
\end{minipage}

\end{center}
\caption{Profiles of 
neutral pions of energies 10~GeV (a-b), 50~GeV (c-d), and 100~GeV (e-f) decaying to two photons. 
Left: Longitudinal shower profiles 
and the cluster centers found by the FCM algorithm on the 4th, 8th and 12th layers. 
Right: the lateral shower profiles of the 8th layer and the clusters
found by the FCM algorithm.} 
\label{fig:twoclusters}
\end{figure}

In order to demonstrate the practicability of the FCM algorithm described in Section~\ref{sect:FCM}, 
data for a single $\pi^0$ decay into two photon clusters at different
energies was generated.
Figure~\ref{fig:twoclusters} displays the shower profiles of photons
emitted from $\pi^0$ at three different energies: 10~GeV, 50~GeV, and 100~GeV.
The left column shows the event display of the three cases. 
The two photon showers 
move closer to one another with the increase of $\pi^0$ energy, 
thereby causing an overlap of clusters.  
The FCM algorithm was used to cluster the data points on each plane of
the detector, taking the $x$ -position, $y$ - position and the energy
depositions of each pad into account. 
The fuzzy factor $m$ was varied, and a value of 1.8 was found to be
suitable in resolving the individual clusters.
This value has been also been recommended in \cite{SPalViyogiFCM}.
For all FCM clustering performed, the following parameter values were taken:
\begin{center}
$m =1.8$  ~~~~~ ~~~~~{\rm and}~~~~~~~~~~~ $\epsilon = 0.01$.
\end{center}

\begin{figure}[htbp]
\begin{center}
\begin{minipage}{500pt}
\begin{tabular}{cc}
\includegraphics[scale=.4]{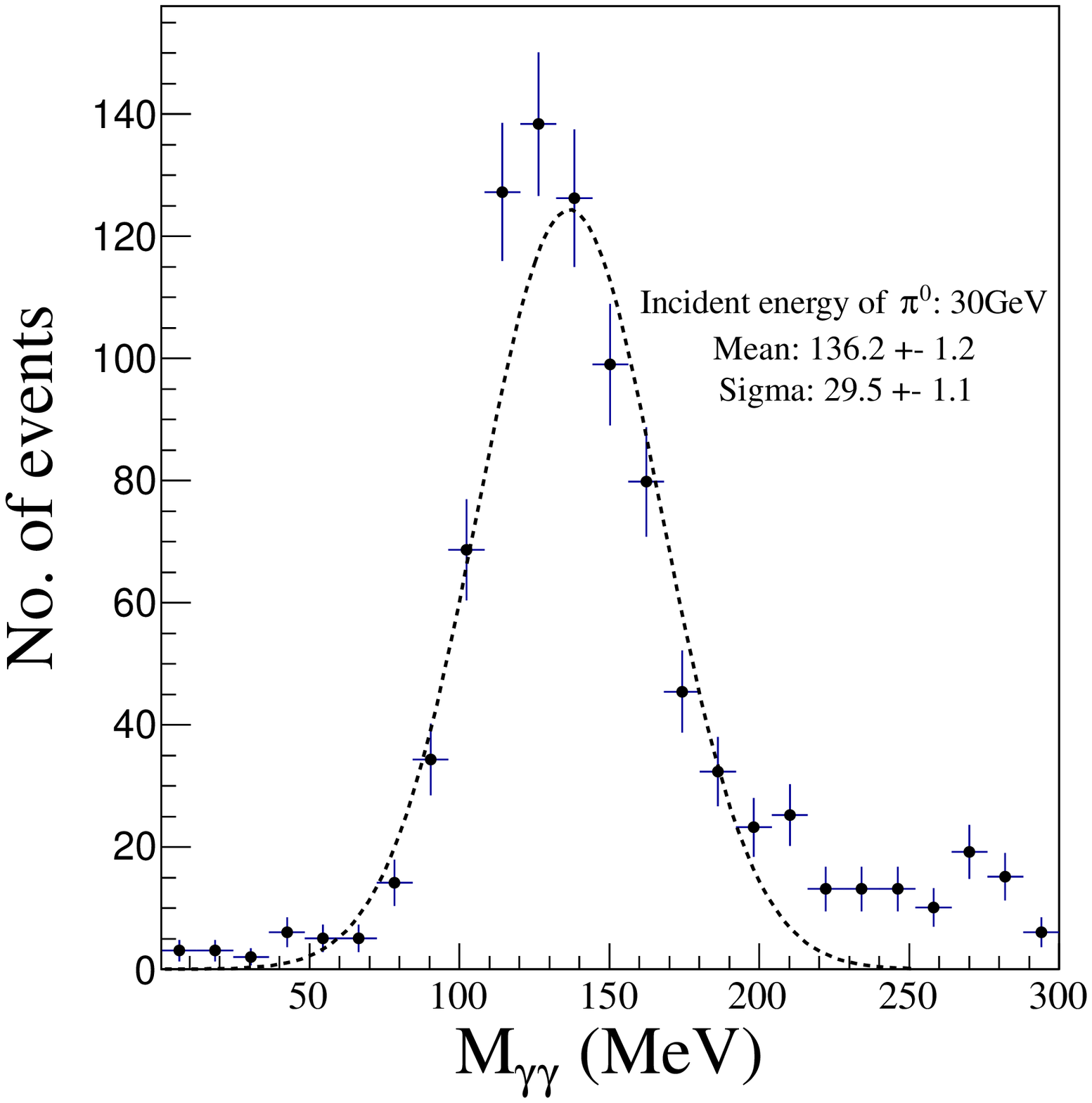}&
\includegraphics[scale=.4]{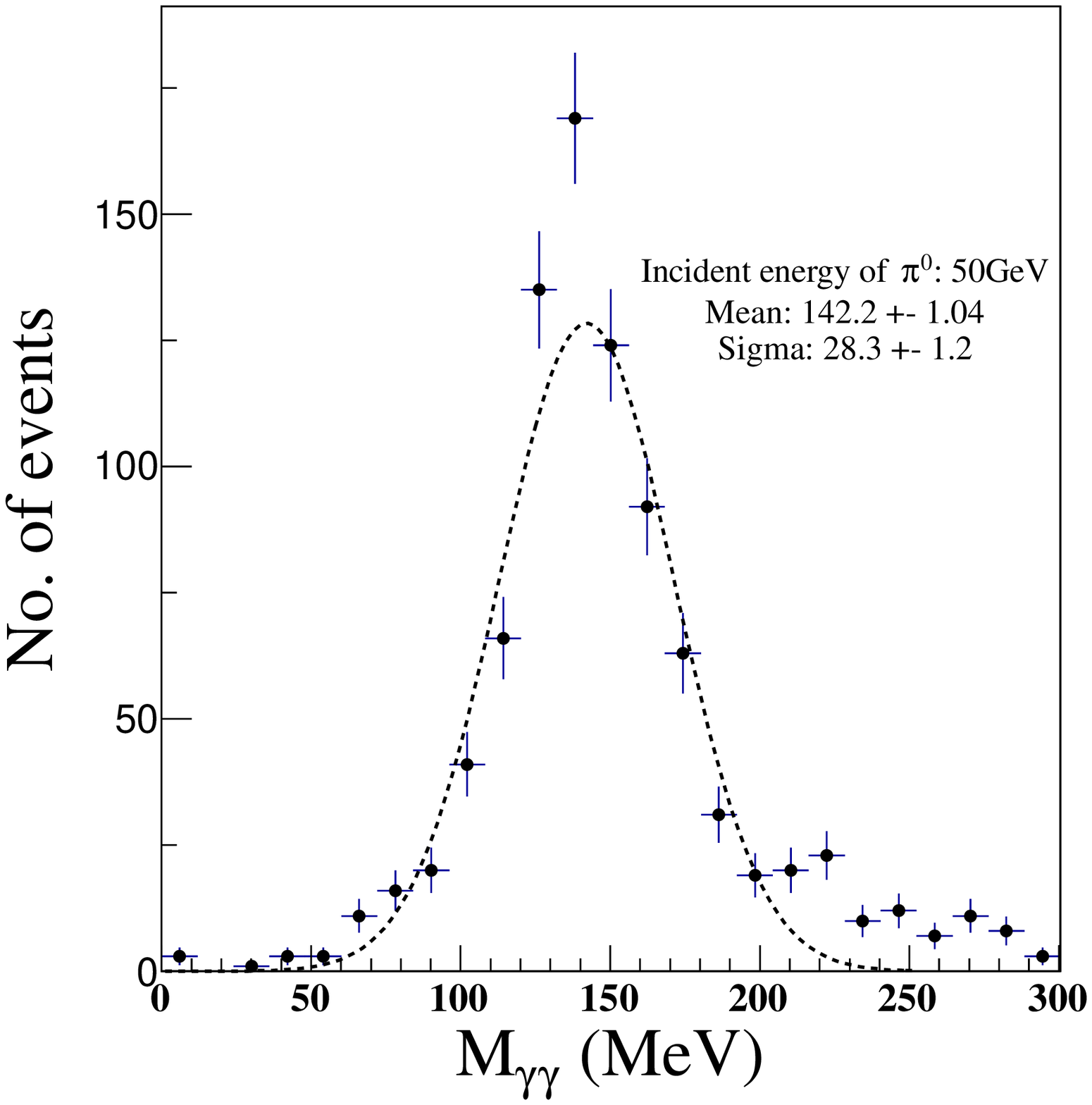} \\
(a) & (b)
\end{tabular}
\end{minipage}

\begin{minipage}{500pt}
\begin{tabular}{cc}
\includegraphics[scale=.4]{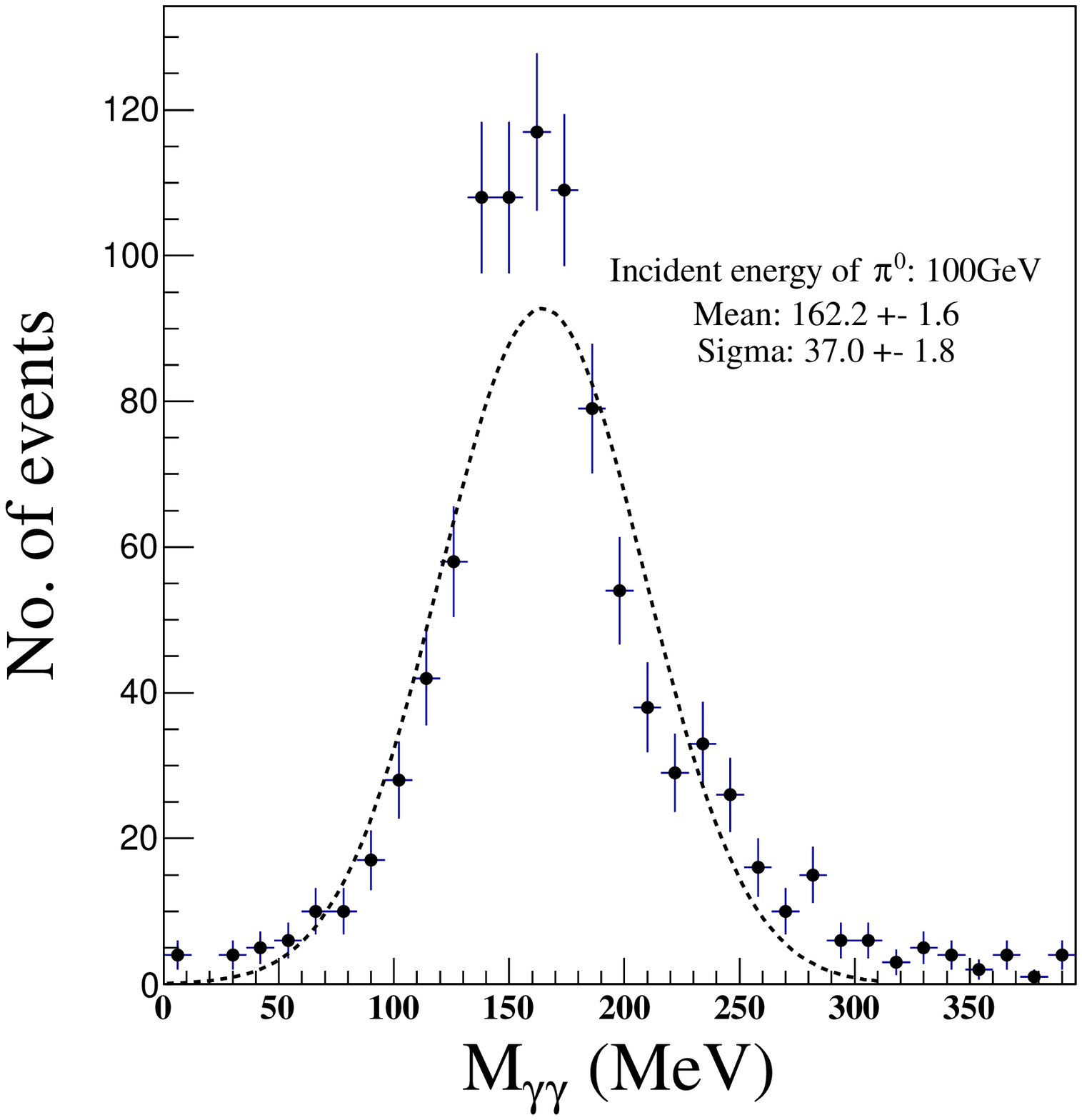}&
\includegraphics[scale=.41]{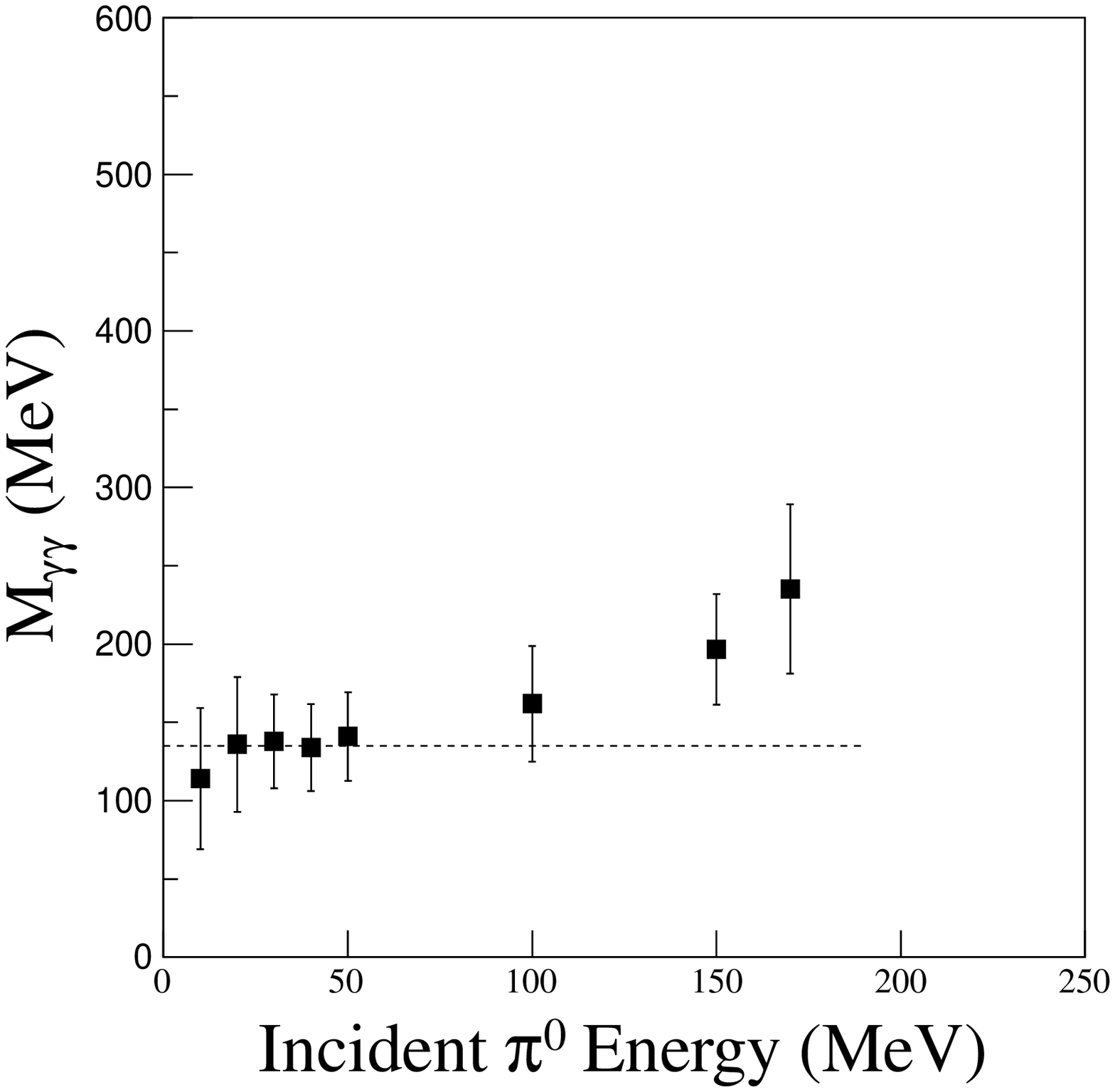}\\
(c) & (d)
\end{tabular}
\end{minipage}
\caption{
Invariant mass distributions of two photons in the
sampling calorimeter using FCM clustering routine for $\pi^0$ energies
of: (a) 30 GeV, (b) 50 GeV, (c) 100 GeV, and
(d) Reconstructed $\pi^0$ mass by the measurement of cluster positions
using FCM. Statistical errors are indicated in the figure. 
}
\label{fig:pi0mass_reconstruction}
\end{center}
\end{figure}

The results of the clustering for layer 8 in terms of lateral
shower profiles are shown in the right column of Figure~\ref{fig:twoclusters}.
The solid dots in the left column indicate
the cluster centers found on each of the three planes. 
The lines represent the photon tracks. It is seen that despite the large extent of overlap, 
the photon clusters are clearly identified by the FCM algorithm, and
the photon paths are successfully tracked using the three layers.
The cluster positions, the tracks and the energy depositions of the
photons have been
used to reconstruct the mass of the $\pi^0$. This is a test of the
working of the reconstruction algorithm, where clustering of the hit
pads plays an important role.
Figure~\ref{fig:pi0mass_reconstruction}(a-c) shows the invariant mass reconstruction
using the FCM algorithm, which essentially indicates the quality of
$\pi^0$ reconsturction for three different energies.
Figure~\ref{fig:pi0mass_reconstruction}(d) shows the $\pi^0$ mass, reconstructed from
the decay of photons for $\pi^0$ of different energies. The
statistical errors indicated on this figure are the {\it rms} values
of the invariant mass distributions. The granularity of the detector
and the limitations of the clustering routine limit reconstruction of
$\pi^0$. As seen from Figure~\ref{fig:pi0mass_reconstruction}(d), up
to 100 GeV energy, there is a reasomable mass reconstruction, beyond
which it deviates.

\section{Application of dFCM to Calorimetric Data}
\label{sect:dFCMapplication}

The dFCM clustering technique as discussed in Section~\ref{sect:dFCM}
works equally well for the calorimetric data discussed
in the previous section. 
Since FCM is a static algorithm that takes all data points into account
at one time, it is sometimes unable to identify clusters with
non-uniform data patterns, like those encountered in a real experimental scenario
where a variety of particles hit the calorimeter. The dynamic version of the FCM algorithm works better to resolve very close clusters.

\begin{figure}[htbp]
\begin{center}
\begin{tabular}{cc}
\includegraphics[scale=.25]{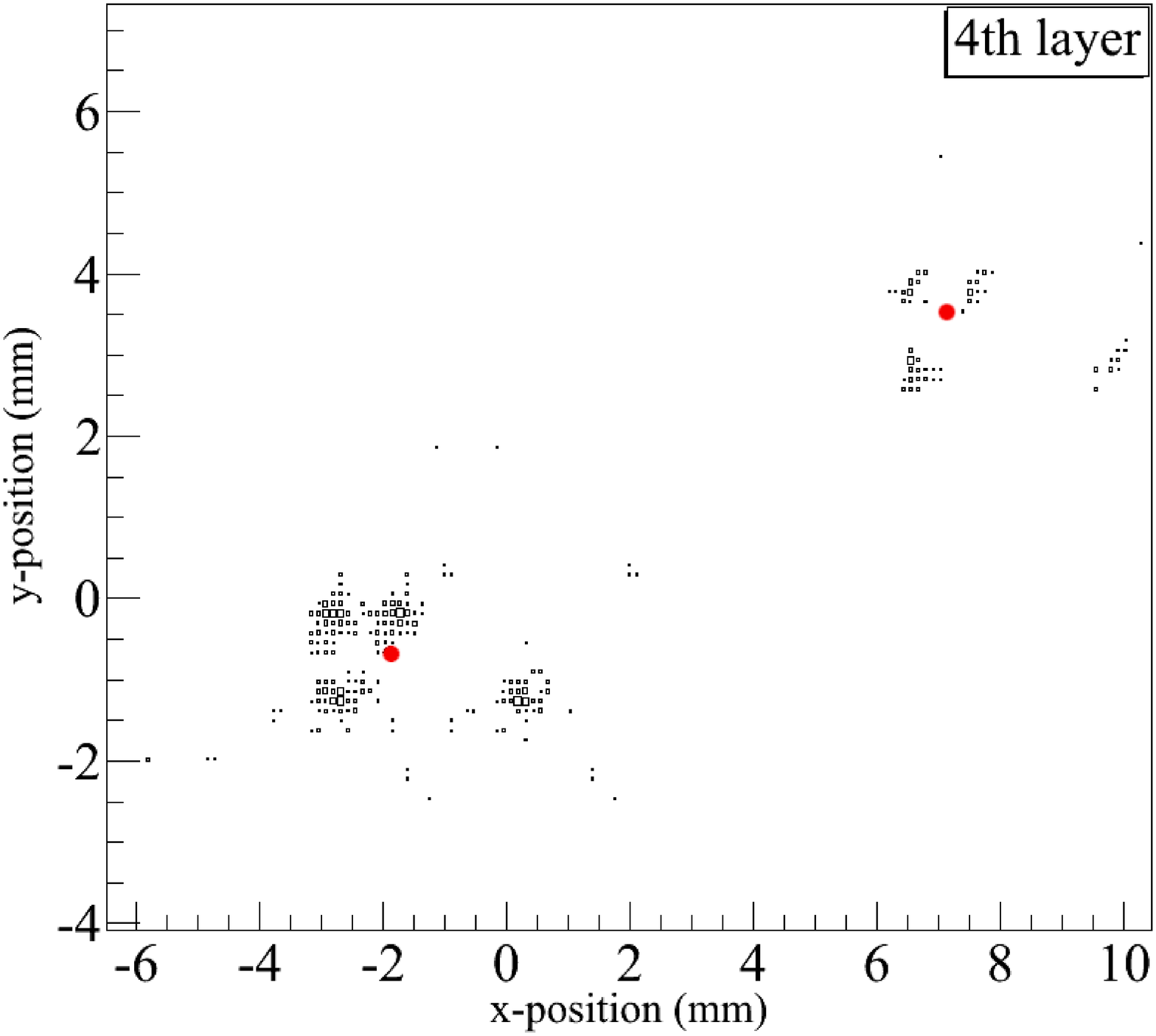} &
\includegraphics[scale=.25]{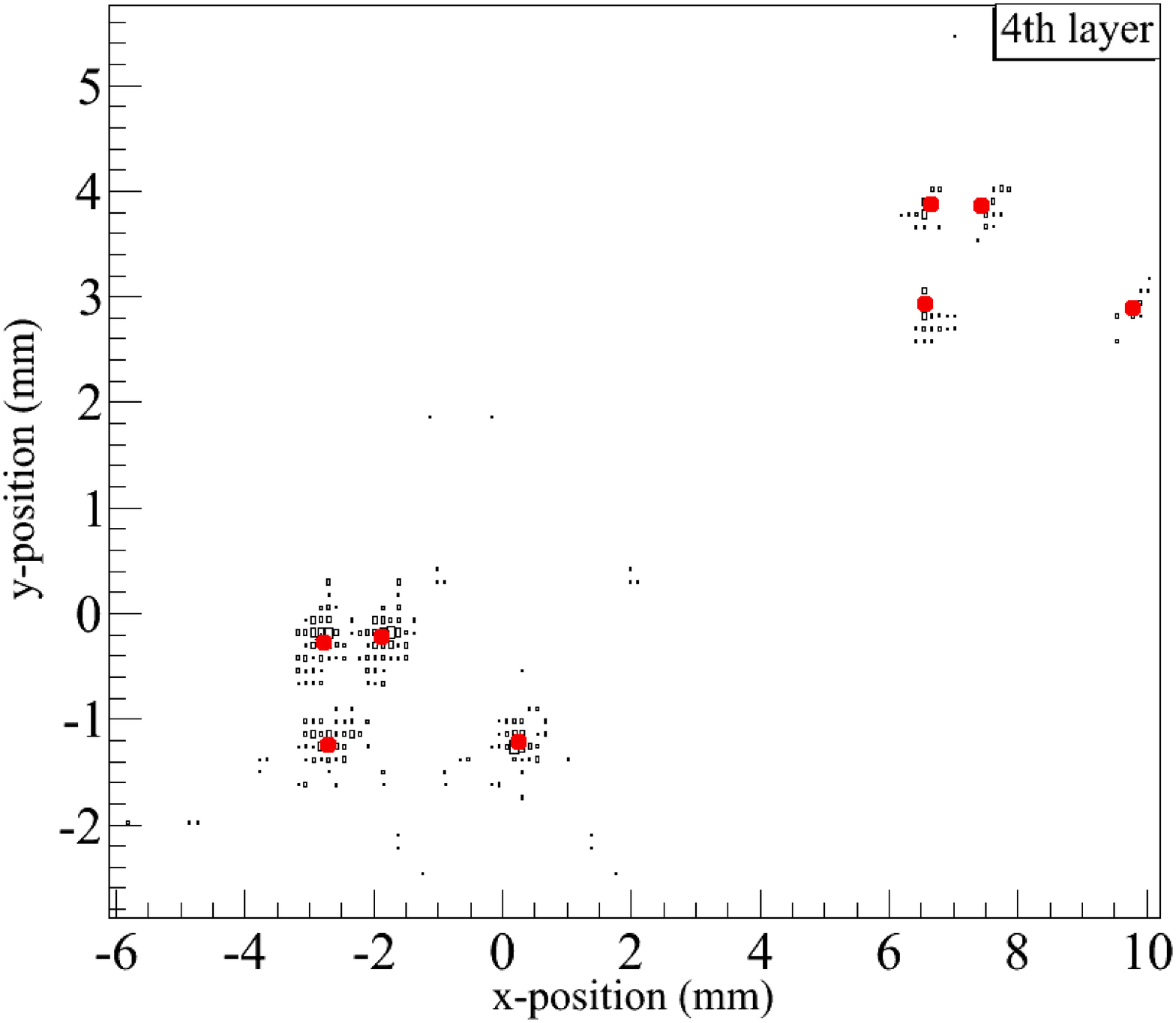} \\
(a) & (b)
\end{tabular}
\caption{(a) Raw data points from  the fourth layer of the calorimeter
and  the clusters found by using the simple FCM algorithm. 
The minimum of Xie-Beni validity index was obtained for the two cluster centers shown. 
(b) The results of clustering obtained by using the dFCM algorithm. All the clusters are properly
accounted for.}
\label{fig:fcm8clusters}
\end{center}
\end{figure}

In order to determine the spatial resolving power of both FCM and dFCM, a sample of eight clusters on the fourth layer of the
calorimeter was generated. Figure~\ref{fig:fcm8clusters} shows the
profiles of the eight clusters. Each data point represents the
position and energy deposition on each of the small pads 
(0.1~cm~$\times$~0.1~cm).
First, the FCM clustering technique was performed, with an input
cluster range of 2 to 10.
The Xie-Beni index was used to select the best set of cluster centers.
The minimum value of the index was obtained for two cluster centers, as the
data appears to be distributed in two major groups of four clusters each. 
The solid points of the left panel of Figure~\ref{fig:fcm8clusters} shows the two cluster centers obtained using the FCM algorithm and the Xie-Beni index on the raw data.
The FCM clustering could not resolve the individual clusters, as they appear to be within a group.

\begin{figure}[hpbt]
\begin{center}
\includegraphics[scale=.37]{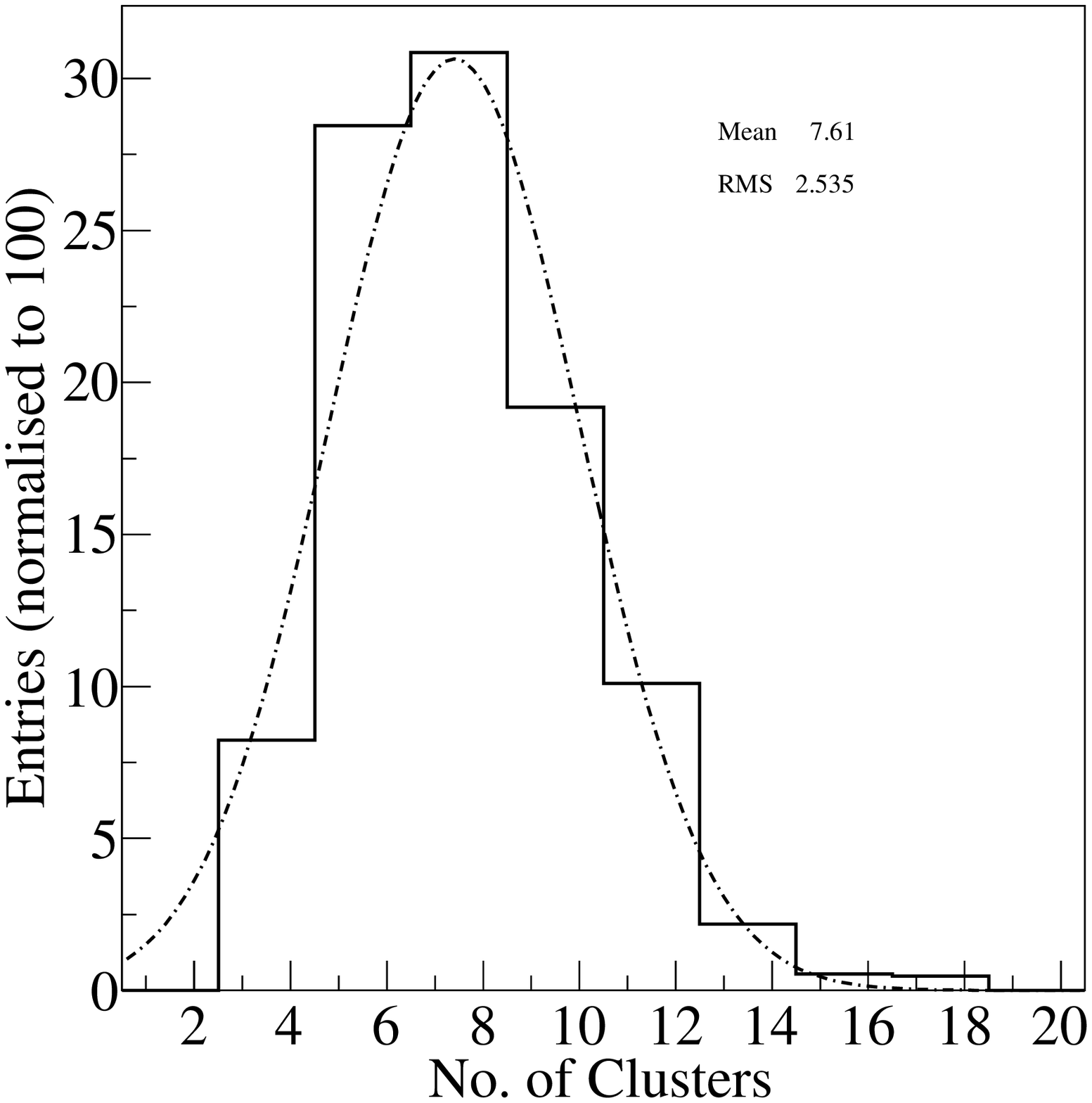}
\caption{
Number of clusters identified by the 
dFCM algorithm for known sets of eight clusters. The histogram is
normalized to 100 clusters. A peak is observed around 8 clusters, and within
$\pm 1$, a total of 80 clusters can be accounted for. 
}
\label{fig:8clusters}
\end{center}
\end{figure}

Next, the dFCM algorithm was applied to the same data, and the Xie-Beni
index was used to find the number of clusters. 
For all dFCM clustering performed, the following parameter values were taken:
\begin{center}
$m =1.8$, ~~~~~ $\epsilon = 0.01$, ~~~~~~ $\mu = 0.8$, ~~~~~ 
  {\rm Number of initial points =}  10.
\end{center}
The right panel of Figure~\ref{fig:fcm8clusters} shows the result of
clustering using the dFCM algorithm. As can be seen, there are some
isolated scattered data points that affected the number of clusters
obtained. The energy depositions of the scattered data points were
low, and below that of one MIP (minimum ionizing particle). Thus, a cut of
one MIP on a cell level was applied so that scattered points were not
considered before clustering. 
With this cut, we observe that the dFCM method was able to resolve and
identify all the eight individual clusters.
This was possible because the dFCM method, as was emphasized earlier,  
allows clustering to be performed on various
stages of data as the data points 
stream in, and adapts to the structure of data at each instant.
The dFCM algorithm is more suitable to resolve clusters that
are not uniformly distributed. 

In order to demonstrate the power of
the dFCM algorithm, around 1000 events of eight clusters were synthesized with varying strengths as well as positions. For
each event, the dFCM algorithm was applied. In Figure~\ref{fig:8clusters}, the 
output of the
dFCM clustering in terms of number of identified clusters has been
plotted for the case of eight known clusters. The histrogram is
normalized to 100. 
A peak is observed around 8 clusters, and within
$\pm 1$, a total of 80 clusters can be accounted for.
That means for known 8 clusters, about 80\% of the clusters
 can be found with number of clusters between 7-9. 
This tells about the quality of the clustering.

{\it A note on the validity index: } In order for a validity index to accurately select a set of cluster centers that represent a set of clusters well, a clustering algorithm must be capable of finding the clusters in the first place. That is to say, the clustering algorithm must find cluster center coordinates in such a way that minimizes the validity index that will ultimately be used to select the best set of coordinates.   Even though the FCM algorithm performed clustering from 2 to 10 cluster centers, the coordinates that it found when it searched for 8 cluster centers did not eventually yield the minimal value of the Xie-Beni index.  The index selected the better representative of the data to be 2 cluster centers.  However, when the same index was used to select the best set of clusters found by dFCM, the minimal value was obtained when all 8 individual clusters were resolved.

While it is true that validity indices are by no means generic, it is safe to say that dFCM can at least find suitable clusters centers when searching over a range of them, whether or not the validity index can identify them.  This is also true for FCM, which can be forced to find a certain set of clusters by selecting a single number of clusters to find.  Keeping the role of the validity index in mind, the same experiment was repeated with the partition coefficient described by Equation~\ref{eq:PC} and its modification, described by Equation~\ref{eq:MPC}.  With FCM, the two coefficients were maximized when 2 cluster centers were obtained, whereas the PC selected 7 cluster centers and the MPC 9 cluster centers, with dFCM. Even when the cluster center ranges were modified to [3,10], FCM did not provide cluster centers that resolved the clusters well, and the validity indices indicated that having only 3 cluster centers was the better option.

\begin{figure}[htbp]
\centering
\begin{tabular}{cc}
\includegraphics[scale=.40]{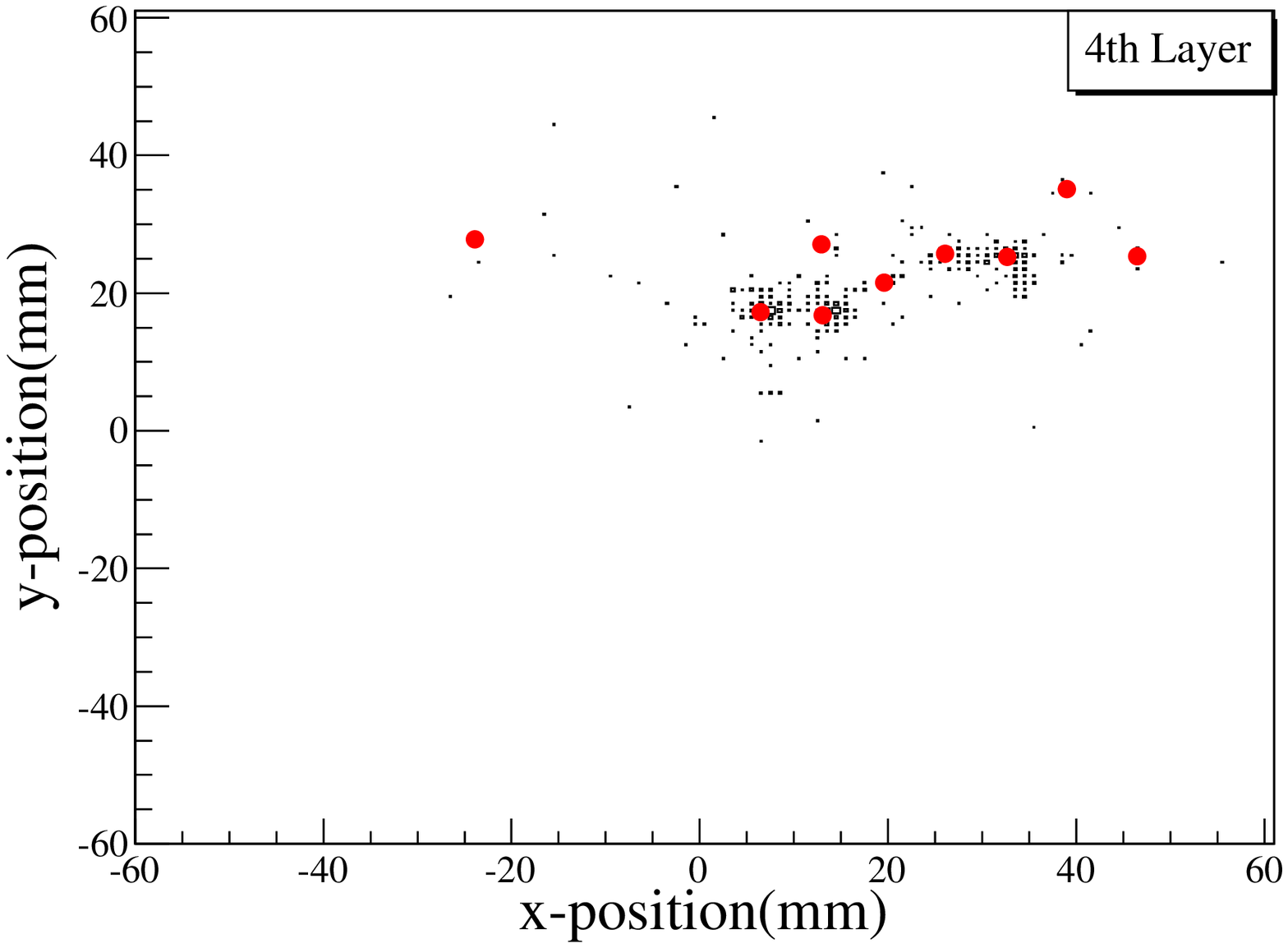} &
\includegraphics[scale=.40]{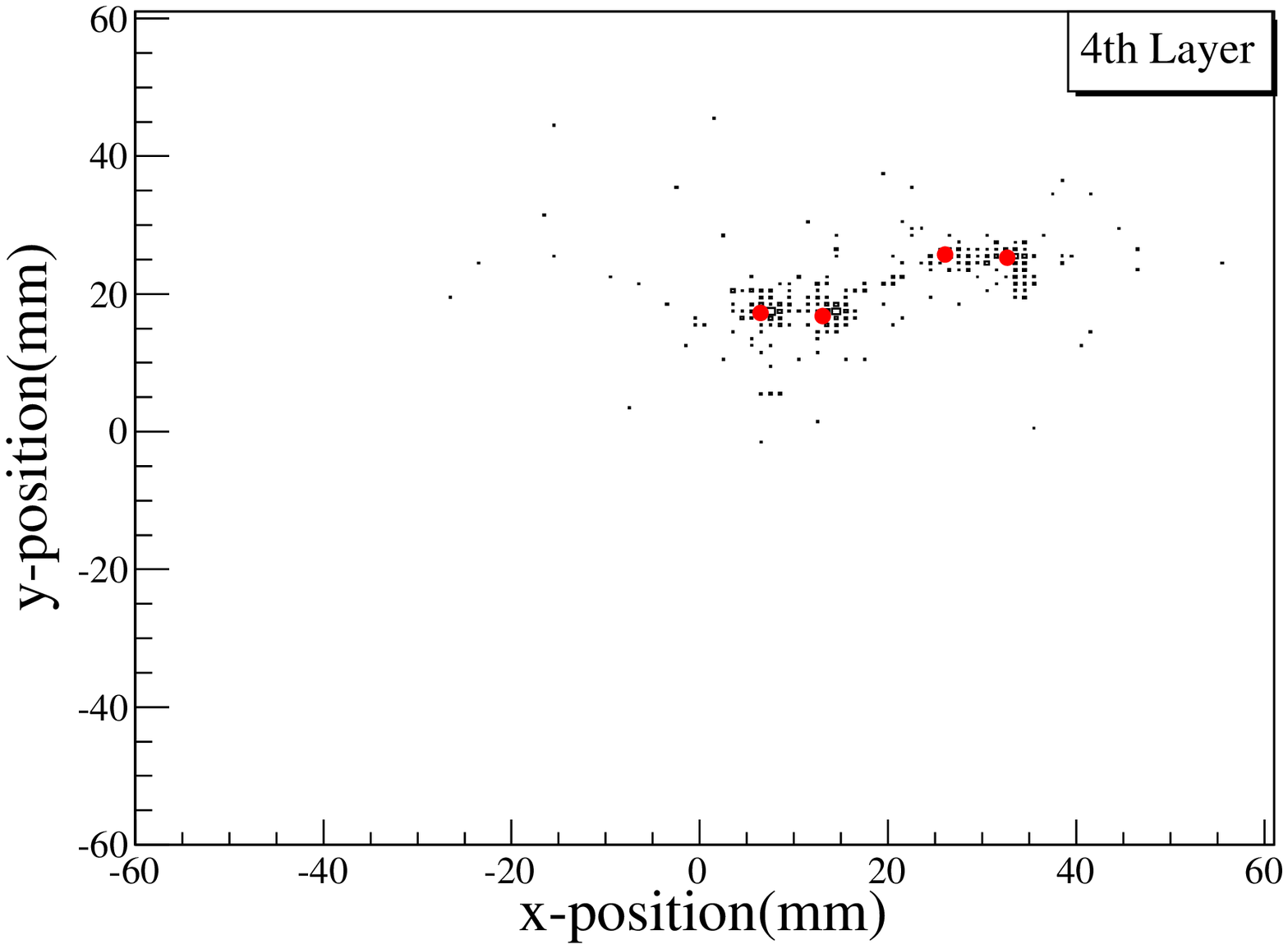} \\
(a) & (b) \\
\includegraphics[scale=.40]{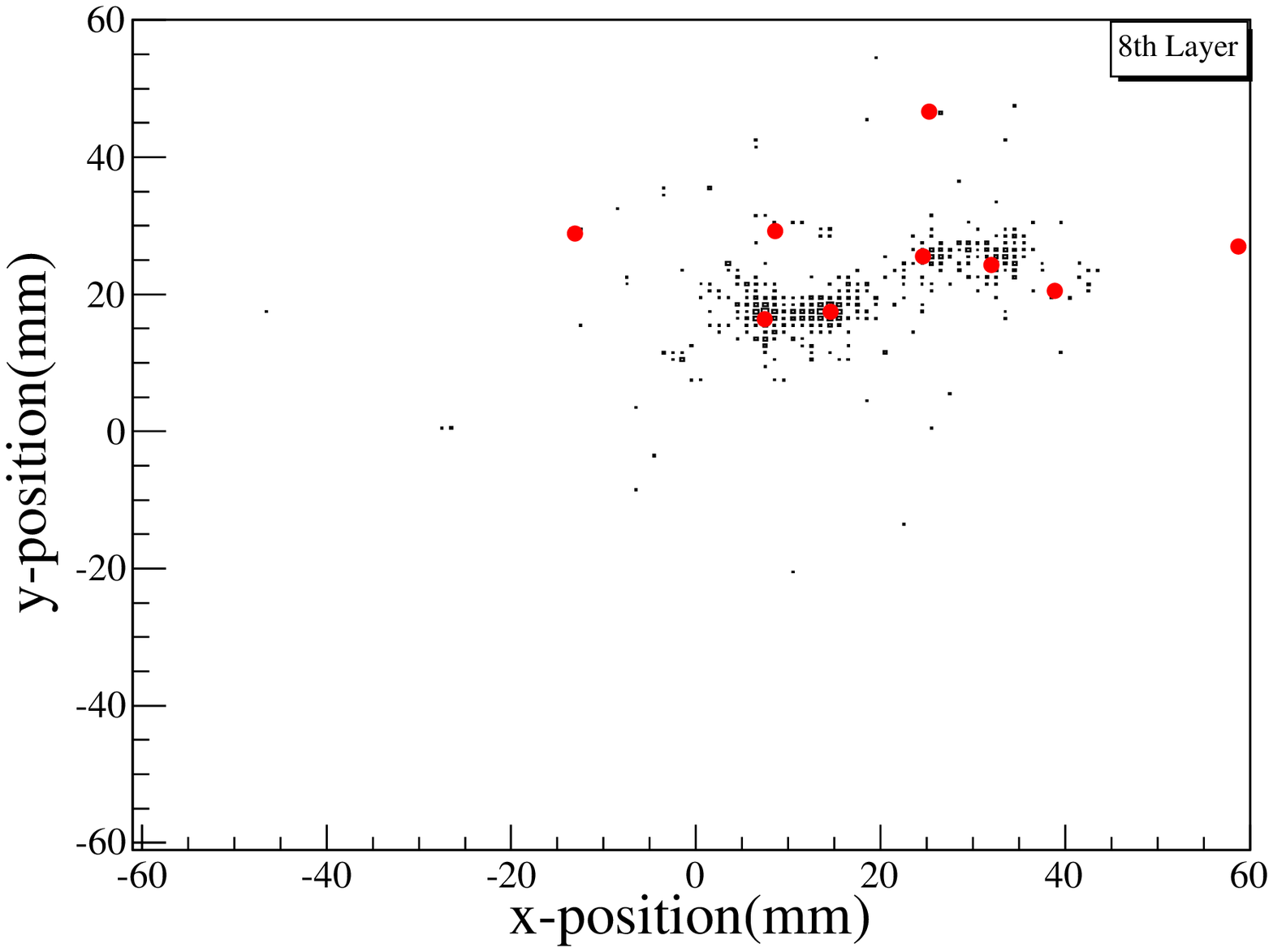} &
\includegraphics[scale=.40]{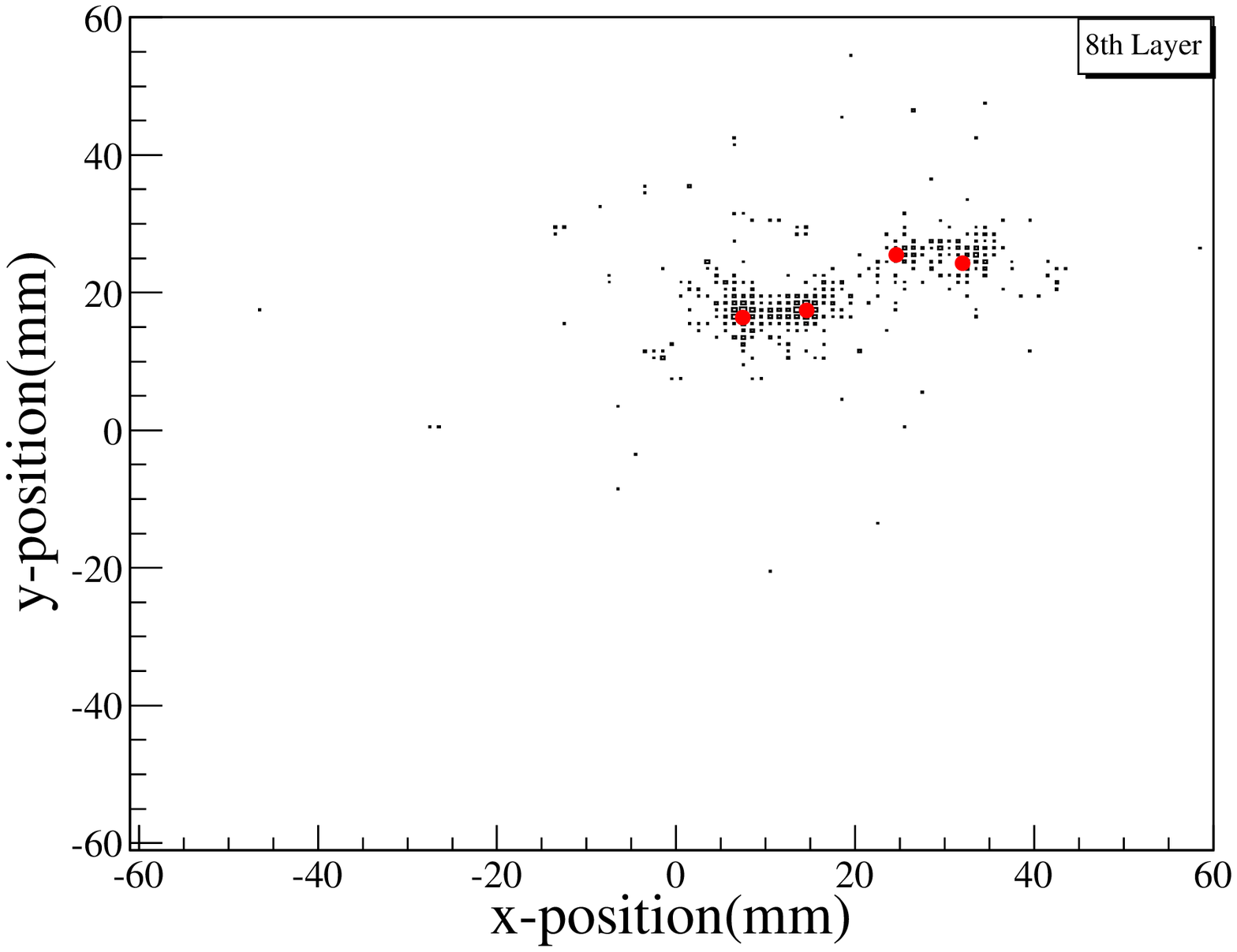} \\
(c) & (d) \\
\includegraphics[scale=.40]{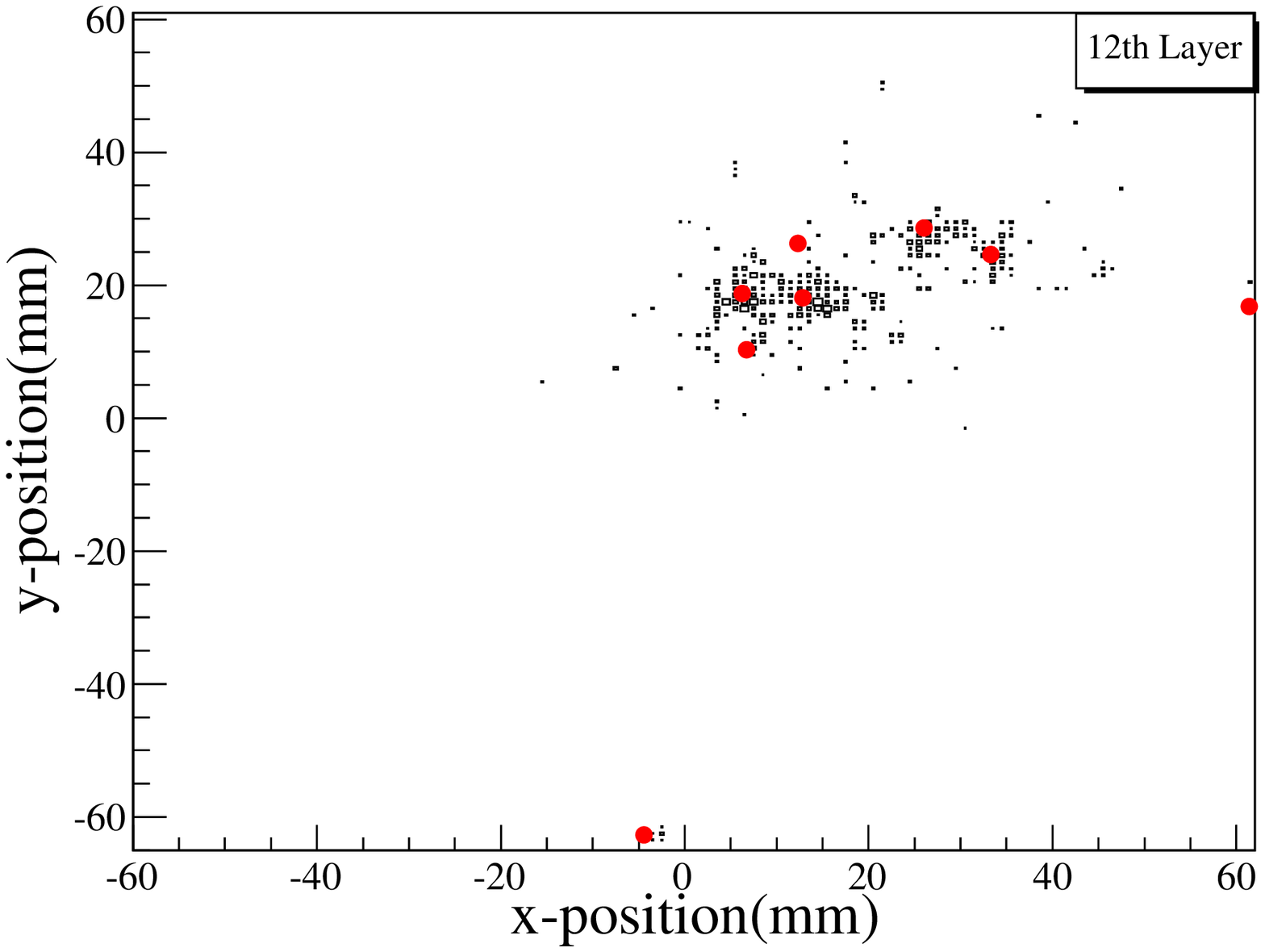} &
\includegraphics[scale=.40]{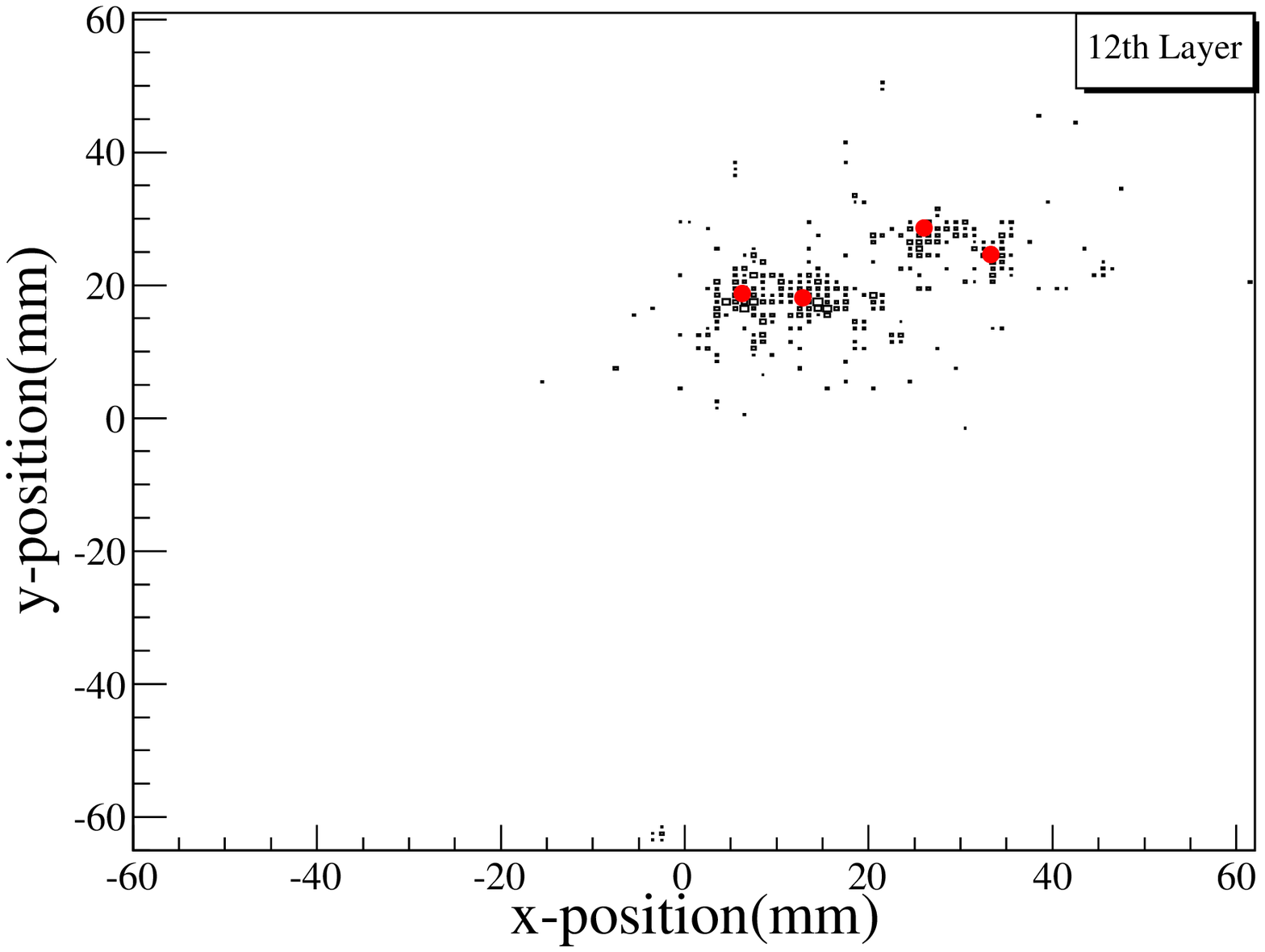} \\
(e) & (f) 
\end{tabular}
\caption{Hits on 4th, 8th and 12th layer of the calorimeter, obtained
from two closely placed neutral pions, each of 50~GeV energy.
The left panels, (a), (c), and (e) show the results of the dFCM clustering
routine, shown by the large solid points. The right panels, (b), (d)
and (f) show the clusters which survive after a cut on the number of
cell hits in a cluster.
}
\label{fig:lastfigure}
\end{figure}

Finally, as an application of the clustering algorithm to
calorimetric data, a simulation of two 50 GeV neutral pions, placed
close together, were
studied for the layers 4, 8 and 12 where higher resolution
silicon pads are placed.  The data points in 
Figure~\ref{fig:lastfigure} show the hits on the 4th, 8th and 12th
layers of the calorimeter. The four clusters can be seen to be
reasonably close together. The 
dFCM clustering algorithm was carried out to identify the clusters. 
A cut off of 89 keV, corresponding to
one MIP, was applied before the clustering.  
The left side panels of (a), (c) and (e) give the results of the
clustering routine with only a threshold of one MIP, applied before clustering.
It is seen that in all cases, the major clusters have been accounted for. Some
scattered data points that have higher energy depositions give
additional clusters, which are essentially outliers.
It is found that the energy depositions and number of hit cells of the
outlier clusters are much smaller compared to the 
four main clusters for each layer.  
Using these data sets, the behaviour of the outliers are studied. It
turns out that most of the clusters in the outliers are made up of
a few cells with low values of deposited energies. 
Thresholds on the number of cells per cluster and deposited energy
can be used to eliminate the outlier clusters. The  panels, (b), (d),
and (f) on the
right  side of the figure show the results of clustering after the application of
thresholds on the cell hits, which eliminate the outliers in each of
the layers. In case of actual calorimetric data, the thresholds can be
optimized and applied in order to obtain the photon clusters.

\section{Summary}
\label{sect:summary}

In order to reconstruct the physical information extracted from
calorimeters and other particle detectors, a suitable clustering technique must be
applied.  The ability of fuzzy clustering
 to assign data points to more than one
cluster at a time makes it a powerful tool when clusters overlap, as is 
often the case with large particle densities of high energy
experiments. The FCM technique and its dynamic version, dFCM, have
been evaluated to be used in calorimetric data reconstruction. 

A tungsten-silicon sampling calorimeter of 20 radiation lengths was configured using
the GEANT4 package. The calorimeter consisted of 20 layers, each with
one radiation thick tungsten and 300 micron silicon pad detectors.
Layers 4, 8 and 12 consisted of high resolution silicon pads, 
 with 0.1~cm~$\times$~0.1~cm sized pixels. 
The remaining layers were silicon
pad layers with cell dimensions of 1~cm~$\times$~1~cm. 
The fuzzy c-means (FCM) clustering technique was applied 
to all the layers for the two photon clusters obtained 
after the decay of a neutral pion. The study demonstrated 
how the clusters and photon tracks could be identified for 
different energies of neutral pions.
The tracks are successfully
identified by locating the clusters in the 3 high resolution pad
layers, and the total energy of a given track is obtained by summing
over clusters of all pad layers belonging to the track. This provides a method
for successful $\pi^0$ reconstruction in the calorimeter.

The dynamic version of the FCM algorithm (dFCM) 
allows clusters to be generated and eliminated as required
 as data streams in. The dFCM algorithm successfully identified 
the individual clusters in a non-uniformly distributed set of 8 photon
clusters, whereas FCM was shown to consider a group of clusters 
to be one large cluster. The dFCM technique was also applied to 
a simulation of two 50 GeV neutral pions decaying to two photons 
each. Each of the four photons were easily identified, despite the 
presence of small clusters of outliers. The energies and the number 
of points in each of the outlier clusters are usually small enough 
for a suitable elimination criteria to be defined. However, as 
different energies as well as different layers give rise to varying
 densities and different cluster structures, a meticulous study needs
 to be performed in order to come up with proper discrimination
 criteria. The discrimination criteria will have to be tuned in a
 realistic environment with different colliding systems and colliding
 energies.

\end{document}